\documentclass[preprint, pre, amsmath, amssymb, aps, superscriptaddress]{revtex4-2}

\usepackage{graphicx}
\usepackage{dcolumn}
\usepackage{bm}
\usepackage{comment}
\usepackage{enumerate}
\usepackage{amsmath}
\usepackage{xcolor}

\usepackage{hyperref}
\hypersetup{hidelinks}
\usepackage{footnotebackref}

\begin{document}

\title{Analyzing the effect of cell rearrangement on Delta-Notch pattern formation}

\author{Toshiki Oguma}
\affiliation{Kyushu University Graduate School of Medical Sciences, Fukuoka 812-8582, Japan}
\author{Hisako Takigawa-Imamura}
\affiliation{Kyushu University Graduate School of Medical Sciences, Fukuoka 812-8582, Japan}
\author{Tomoyasu Shinoda}
\affiliation{Nagoya University Graduate School of Medicine, Nagoya 466-8550, Japan}
\author{Shuntaro Ogura}
\affiliation{Nagoya City University Graduate School of Medical Sciences, Nagoya 467-8601, Japan}
\author{Akiyoshi Uemura}
\affiliation{Nagoya City University Graduate School of Medical Sciences, Nagoya 467-8601, Japan}
\author{Takaki Miyata}
\affiliation{Nagoya University Graduate School of Medicine, Nagoya 466-8550, Japan}
\author{Philip K. Maini}
\affiliation{Wolfson Centre for Mathematical Biology, Mathematical Institute, Oxford OX2 6GG, UK}
\author{Takashi Miura}
\affiliation{Kyushu University Graduate School of Medical Sciences, Fukuoka 812-8582, Japan}

\date{\today}

\begin{abstract}
    The Delta-Notch system plays a vital role in a number of areas in biology and typically forms a salt and pepper pattern in which cells strongly expressing Delta and cells strongly expressing Notch are alternately aligned via lateral inhibition.
    Although the spatial arrangement of the cells is important to the Delta-Notch pattern, the effect of cell rearrangement is not often considered.
    In this study, we provide a framework to analytically evaluate the effect of cell mixing and proliferation on Delta-Notch pattern formation in one spatial dimension.
    We model cell rearrangement events by a Poisson process and analyze the model while preserving the discrete properties of the spatial structure.
    We find that the homogeneous expression pattern is stabilized if the frequency of cell rearrangement events is sufficiently large.
    We analytically obtain the critical frequencies of the cell rearrangement events where the decrease of the pattern amplitude as a result of cell rearrangement is balanced by the increase in amplitude due to the Delta-Notch interaction dynamics.
    Our theoretical results are qualitatively consistent with experimental results, supporting the notion that the heterogeneity of expression patterns is inversely correlated with cell rearrangement \textit{in vivo}.
    Our framework, while applied here to the specific case of the Delta-Notch system, is applicable more widely to other pattern formation mechanisms.
\end{abstract}

\maketitle

\section{Introduction}
\label{Introduction} 

The Delta-Notch system is a well-studied cell-cell communication system that generates cellular scale periodic patterns \cite{mohr_character_1919, collier_pattern_1996, sprinzak_cis-interactions_2010, vilas-boas_novel_2011, shaya_notch_2011, matsuda_synthetic_2012, uriu_determining_2017, herman_novel_2018}. 
Delta and Notch are, respectively, cell surface ligands and receptors.
After receptor-ligand binding with Delta, the intracellular domain of Notch is cleaved, transported to the nucleus, enhances Notch expression, and suppresses Delta expression.
In this way, Delta inhibits the expression of Delta in adjacent cells.
This process is termed ``lateral inhibition". 
As a result, cells strongly expressing Delta and cells strongly expressing Notch are aligned alternately (the so-called, ``salt and pepper'' pattern) \cite{shaya_notch_2011}.
The Delta-Notch system contributes to cell fate determination in many developmental processes, such as neuroendocrine cell differentiation in the lung \cite{Morimoto_Different_2012}, outer hair cell differentiation in the inner ear \cite{eddison_notch_2000, chrysostomou_delta-like_2012, neves_patterning_2013}, and angiogenesis \cite{claxton_periodic_2004, hofmann_notch_2006, hasan_endothelial_2017, pitulescu_dll4_2017}.
As such, Delta-Notch pattern formation plays a critical role in many developmental processes.

Collier \textit{et al}. constructed the first mathematical model for the Delta-Notch system, which consisted of a spatially discrete ordinary differential equation system which was then analyzed, and necessary and sufficient conditions for a salt and pepper pattern were derived \cite{collier_pattern_1996}.
However, there is stochasticity in the cell-cell interactions and gene expression in signal transduction \cite{cohen_importance_2010, elowitz_stochastic_2002, charlebois_gene_2011}, and a number of subsequent theoretical studies have incorporated stochasticity and revealed that, while low-intensity noise contributes to fine-grained pattern formation, high-intensity noise disrupts the salt and pepper pattern \cite{cohen_importance_2010, rudge_effects_2008, reppas_extrinsic_2016}.

However, little research has been conducted to investigate the effect of positional perturbations arising from cell mixing and proliferation, despite these phenomena being generally observed.
It is known that cells actively move and proliferate during the process of cell fate determination by the Delta-Notch system \cite{noguchi_directed_2015, tsao_epithelial_2016, arima_angiogenic_2011, luo_arterialization_2020, riahi_notch1dll4_2015}.
In the Delta-Notch system, the direct contact of the cell membrane mediates signal transduction.
Therefore, cell rearrangement by cell mixing and proliferation should significantly affect Delta-Notch pattern formation since the cells of interacting neighbors are changing.
Germano \textit{et al.} have used a computational model to show that excessive cell turnover homogenizes Delta expression \cite{germano_mathematical_2021}, while Stepanova \textit{et al.} developed a computational model to investigate how vascular structures are rearranged in response to the VEGF-Delta-Notch system \cite{stepanova_multiscale_2021}.
However, to analytically understand the effect of cell rearrangement on pattern formation, a simpler model is required.

In this study, we provide a framework to analytically evaluate the effect of stochastic and spatial perturbations arising from cell mixing and proliferation.
We construct a simple mathematical model that incorporates Delta-Notch interaction and cell rearrangement events (cell mixing and proliferation) in one spatial dimension.
Our numerical simulations and mathematical analysis show that the effect of cell rearrangement is to stabilize the homogeneous steady state, and we provide a framework to analytically evaluate the stability of the pattern dynamics.
Furthermore, we experimentally confirm our ideas through observations of the murine retinal vasculature.

\section{Materials and Methods}

\subsection{Numerical simulations}
The numerical simulations were performed using Mathematica (Wolfram) and Julia (MIT), and we used periodic boundary conditions and an explicit Euler scheme.

\subsection{Animals}
H2B-mCherry transgenic mice were provided by the Laboratory for Animal Resources and Genetic Engineering, RIKEN Center for Developmental Biology \cite{abe_establishment_2011}.
All animals were handled in accordance with Nagoya University Guidelines on Laboratory Animal Welfare.

\subsection{Organ culture of mouse retina}
Mouse pups at postnatal day 5 (P5) were anesthetized on crushed ice and decapitated. 
Their eyes were enucleated and placed in Hanks' Balanced Salt Solution in a 35-mm Petri dish. Using fine forceps, the periorbital connective tissue was removed from the eyeball. 
We made a small hole in the cornea using a 26G injection needle. 
Starting at the hole in the cornea, the sclera, choroid, and retinal pigment epithelia were peeled away. 
The retina was isolated from its anterior segment using microscissors \cite{ogilvie_reliable_1999} and cut into small pieces to prevent focus drift caused by retinal plane curvature. 
The retina was embedded in collagen gel solution ($500$ {\textmu}l) and 1/1000 IB4-Alexa on a 35-mm Petri dish and incubated for 30 min at room temperature to allow the collagen gel to solidify. 
The dish was incubated for 30 min at $37^{\circ}$C to further solidify the collagen gel. 
We added 2 ml of medium [DMEM/F-12 + 10\% FBS + 1/5000 isolectin B4 (IB4)-Alexa488 + 500 ng/ml FGF2] to the dish and visualized the blood vessels.
We performed time-lapse observations using a BX61 W1 upright microscope (Olympus), CSU-X1 (Yokogawa) with iXon+ EMCCD (Andor)($\times$ 20, 5 min/frame, 12 h).
Images were analyzed with Fiji \cite{Schindelin2012}, using the TrackMate plugin \cite{Tinevez_TrackMate_2017} to manually track the centers of cell nuclei.

\subsection{Cell proliferation assay of mouse retinal vasculature}
For the proliferation assay for the endothelial cells in P5 mouse retinal vasculature, $30$ {\textmu}g/g bodyweight of 5-ethynyl-2'-deoxyuridine (EdU; Thermo Fisher Scientific) was intraperitoneally injected 2 h before sacrifice. 
After fixation with 4\% paraformaldehyde in phosphate-buffered saline, the whole-mount retinas were processed with a Click-iT EdU Alexa Fluor 488 Imaging Kit (Thermo Fisher Scientific), in accordance with the manufacturer's instructions, in conjunction with immunohistochemistry using rabbit anti-Ets-related gene-1 (ERG1) monoclonal antibody (Abcam Ab92513) and Cy3-conjugated donkey anti-rabbit IgG secondary antibody (Jackson ImmunoResearch). 
Images were taken using an LSM700 confocal microscope (Zeiss) with ZEN software (Zeiss). We observed five fields of view and performed statistical analysis (Student's t-test) for a fraction of EdU(+) cells in ERG1(+) cells.

\section{Models}
\subsection{Classical Delta-Notch model}
To model the effect of cell mixing or proliferation on Delta-Notch pattern formation, we started with a version of the Collier model \cite{collier_pattern_1996}. 
In this model, the Delta and Notch expression of a cell $x$ (${D}_x$ and ${N}_x$, respectively) in a one-dimensional cell line were modeled (Fig.~\ref{fig:1}(A)) as below:
\begin{eqnarray}\label{eq:Collier}
\frac{d {D}_x}{d t}&=& v \left( \frac{1}{1+ \beta {{N}_x}^h} - {D}_x \right) \nonumber \\
\frac{d {N}_x}{d t}&=& \frac{r \ \left( D_{x-1} + D_{x+1} \right)}{1+ r \ \left( D_{x-1} + D_{x+1} \right)} - {N}_x .
\end{eqnarray}
Here, the parameter $v$ denotes the reaction speed of Delta dynamics relative to that of Notch. The parameters $h$ and $\beta$ denote the Hill coefficient and the intensity of Delta suppression by Notch, respectively.
The parameter $r$ denotes the intensity of Notch activation by the Delta presented in neighboring cells.
The number of cells is $n$ and the position of the cell is $x \ (x \in \mathbb{N}, 1 \le x \le n)$.
We used a one-dimensional model because it is tractable analytically, and the distinct salt and pepper pattern of Delta-Notch expression has been reported in endothelial cells which are aligned one-dimensionally \cite{claxton_periodic_2004, hofmann_notch_2006, herman_novel_2018}.
We assume that the number of cells is sufficiently large so that we can use periodic boundary conditions.
This is because the effect of boundary conditions is confined near the boundary, and the global pattern we focused on is minimally affected by the precise form of the boundary conditions if the system size is large.
We confirmed, using numerical simulation, that the main results of this study are robust to different imposed boundary conditions (results not shown).

In the Collier model we use (\ref{eq:Collier}), whether or not a salt and pepper pattern emerges depends on the model parameters ($v, \beta, h, r$).
The necessary and sufficient conditions for salt and pepper pattern formation are obtained by performing a standard linear stability analysis (Appendix A), requiring that the maximum eigenvalue be greater than zero:
\begin{equation}\label{eq:lambda}
\lambda_{\text{max}} = \frac{-(a+d) + \sqrt{(a+d)^{2}-4(a d - 2b \alpha))}}{2} > 0,
\end{equation}
where $a = v, \ b = (\beta h v (N^{0})^{h-1})/\left(1+\beta (N^{0})^h \right)^{2}, \ d = 1$, $\alpha = {r}/(\left(1+2 r D^{0}\right)^{2})$ and $(D^0,N^0)$ is the spatially homogeneous steady state of the Collier model (\ref{eq:Collier}).
Based on this analysis, we proceeded to investigate how pattern formation is altered by cell mixing and proliferation.

\subsection{Cell mixing model}
To introduce the effect of cell mixing on the Delta-Notch model (\ref{eq:Collier}), we modeled cell mixing as a series of flips between neighboring cells.
We made several assumptions as follows (Fig.~\ref{fig:1}(B)):
\begin{enumerate}[({B}1)]
\item The positions of the neighboring cells are randomly exchanged by cell flips in a single step.
\item Flips occur according to a Poisson process with intensity $p$ in each pair of the cells.
\end{enumerate}

Let the vertical vectors $\boldsymbol{D}$ and $\boldsymbol{N}$, respectively, denote Delta and Notch expression in each cell as below:
\begin{align}
    \boldsymbol{D} &= (D_1, D_2 \cdots, D_x, \cdots, D_n)^{\mathsf{T}} \nonumber \\
    \boldsymbol{N} &= (N_1, N_2, \cdots, N_x, \cdots, N_n)^{\mathsf{T}},
\end{align}
and a flip between cells $j$ and $j+1$ is described by multiplication with the $n \times n$ matrix $A^j$ as below:
\begin{equation}\label{eq:matAj}
    \left\{ A^j \right\}_{k,m} = 
    \left\{
    \begin{array}{ll}
        1      & \text{if } (k = m \text{ and } k \neq j, j+1) \\
               & \text{or } (k = j \text{ and } m = j+1) \\
               & \text{or } (k = j+1 \text{ and } m = j) \\
        0      & \text{otherwise}
    \end{array}
    \right. ,
\end{equation}
where $j+1$ is regarded as $1$ if $j = n$ (periodic boundary condition).
The effect of cell flipping was introduced by stochastically multiplying the matrix $A^j$ by $\boldsymbol{D}$ and $\boldsymbol{N}$.
Hence, our cell mixing model is defined by the system of stochastic differential equations as below:
\begin{align} \label{eq:flipModel}
    d \boldsymbol{D} &= \boldsymbol{f}(\boldsymbol{D}, \boldsymbol{N}) dt + \sum_{j=1}^n (A^j-I) \ \boldsymbol{D} \ dL^{p,j}_t \nonumber \\
    d \boldsymbol{N} &= \boldsymbol{g}(\boldsymbol{D}, \boldsymbol{N}) dt + \sum_{j=1}^n (A^j-I) \ \boldsymbol{N} \ dL^{p,j}_t,
\end{align}
where the functions $\boldsymbol{f}$ and $\boldsymbol{g}$ are the reaction terms of the Collier model (\ref{eq:Collier}), the matrix $I$ denotes the identity matrix and $L^{p,j}_t$ is the Poisson process with intensity $p$, which corresponds to the flip between cells $j$ and $j+1$.

\subsection{Cell proliferation model}
\begin{figure*}[htb]
    \includegraphics[width=17.8cm]{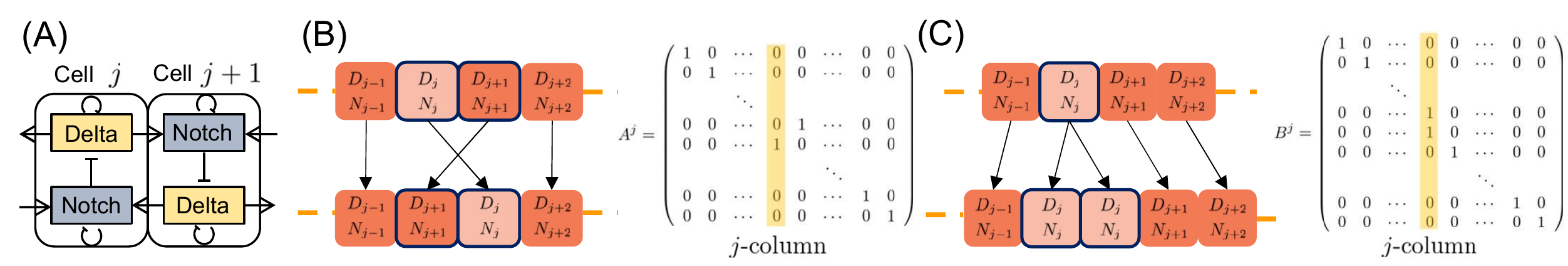}
    \caption{
    (A) Schematic of the Delta-Notch interaction model. 
    Notch expression inhibits Delta expression, Delta expression promotes Notch expression in adjacent cells, and Delta and Notch themselves naturally decay.
    (B) Schematic of the flip event in the cell mixing model and the matrix $A^j$ in (\ref{eq:matAj}).
    The flip event occurs according to the Poisson process with intensity $p$ in each pair of cells.
    (C) Schematic of the duplication event in the cell proliferation model and the matrix $B^j$ in (\ref{eq:matBj}).
    The duplication event occurs according to the Poisson process with intensity $q$ in each cell.
    }
    \label{fig:1}
\end{figure*}

To introduce the effect of cell proliferation on the Delta-Notch model (\ref{eq:Collier}), we modeled cell proliferation as the duplication of a cell.
We also made several assumptions as follows (Fig.~\ref{fig:1}(C)):
\begin{enumerate}[({C}1)]
\item The duplication process occurs in a single step.
\item The new cell is placed to the right of the original cell and inherits the same level of Delta and Notch of the original cell.
\item The duplication process occurs according to the Poisson process with intensity $q$ in each cell \cite{chao_evidence_2019}.
\end{enumerate}

We denote Delta and Notch expression by the vertical vectors $\boldsymbol{D}_n = (D_1,D_2, \cdots, D_n)^{\mathsf{T}}$ and $\boldsymbol{N}_n = (N_1,N_2, \cdots, N_n)^{\mathsf{T}}$, respectively. 
Note that the number of cells (the dimension of the vectors $\boldsymbol{D}_n$ and $\boldsymbol{N}_n$) $n$ increases with time.
Under these assumptions, duplication of cell $j$ is accounted for by defining the $(n+1) \times n$ matrix $B^j$ as below:
\begin{equation}\label{eq:matBj}
    \left\{ B^j \right\}_{k,m} = 
    \left\{
    \begin{array}{ll}
        1      & \text{if } (k = m \text{ and } k \leq j) \\
             & \text{or } (k = m+1 \text{ and } k \geq j) \\
        0      & \text{otherwise}
    \end{array},
    \right.
\end{equation}
and stochastically multiplying this matrix by $\boldsymbol{D}_n$ and $\boldsymbol{N}_n$,  respectively:
\begin{equation}\label{eq:prolModel}
    \begin{aligned}
    &\text { If } d L_{t}^{q, j}=0,
    \left\{\begin{array}{l}
    \boldsymbol{D}_{n}(t+d t)=\boldsymbol{D}_{n}(t)+\boldsymbol{f}(\boldsymbol{D}_n, \boldsymbol{N}_n) d t \\
    \boldsymbol{N}_{n}(t+d t)=\boldsymbol{N}_{n}(t)+\boldsymbol{g}(\boldsymbol{D}_n, \boldsymbol{N}_n) d t
    \end{array}\right. \\
    &\text { If } d L_{t}^{q, j}=1,
    \left\{\begin{array}{l}
    \boldsymbol{D}_{n+1}(t+d t)=B^{j}\left[\boldsymbol{D}_{n}(t)+\boldsymbol{f}(\boldsymbol{D}_n, \boldsymbol{N}_n) d t\right] \\
    \boldsymbol{N}_{n+1}(t+d t)=B^{j}\left[\boldsymbol{D}_{n}(t)+\boldsymbol{f}(\boldsymbol{D}_n, \boldsymbol{N}_n) d t\right]
    \end{array}\right.
    \end{aligned}.
\end{equation}
Note that $n$ will increase with time according to the Poisson process, so the size of $B^j$ will also increase with time.

\section{Results}
\subsection{Numerical simulations with cell rearrangement}
\begin{figure*}[htb]
    \includegraphics[width=17.8cm]{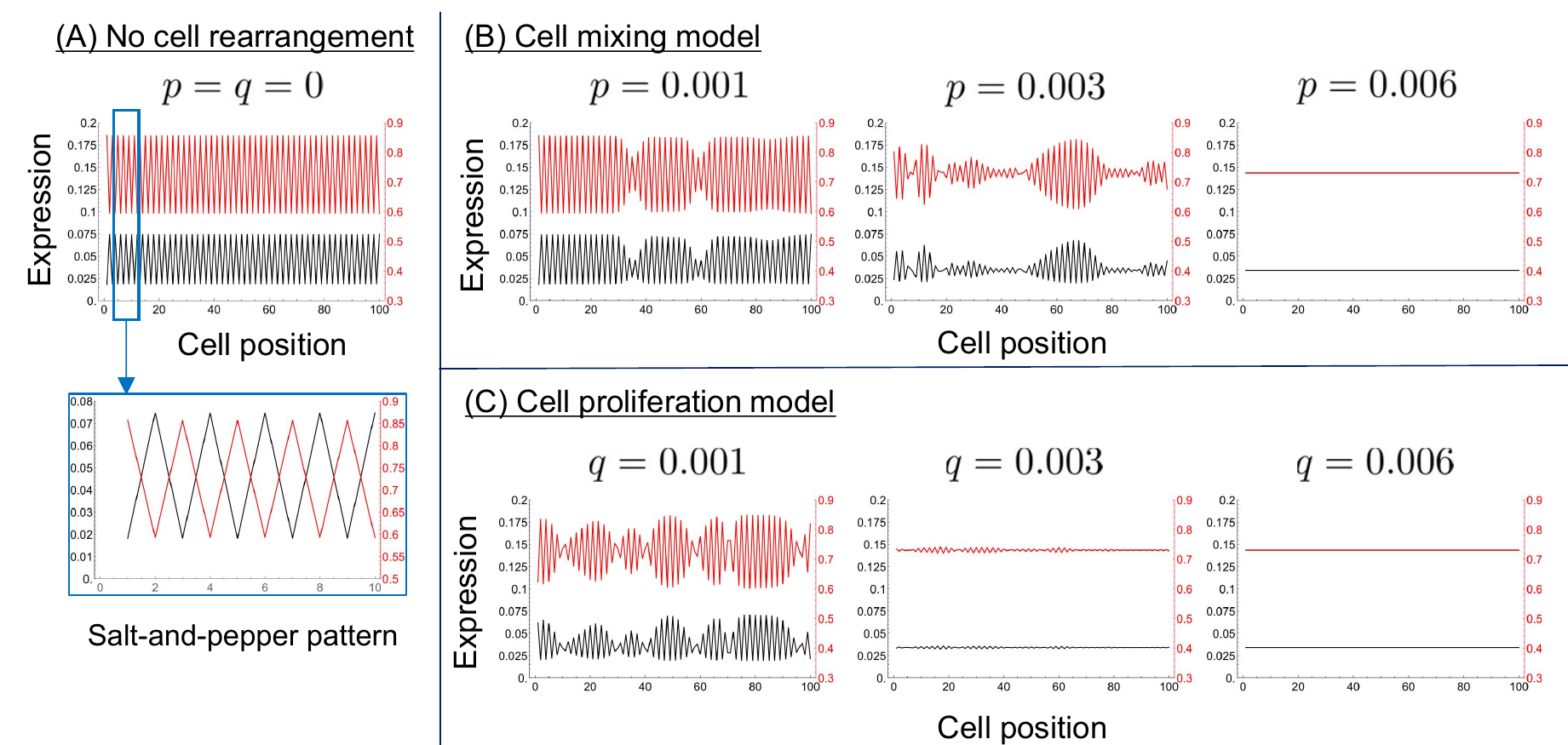}
    \caption{
    Numerical simulations of the standard Delta-Notch model (\ref{eq:Collier}), the cell mixing model (\ref{eq:flipModel}) and the cell proliferation model (\ref{eq:prolModel}).
    (A) Standard model (no cell rearrangement).
    The red line represents Notch expression and the black line represents Delta expression.
    Delta and Notch are alternately expressed, and the classical salt and pepper pattern emerges.
    (B) Cell mixing model (\ref{eq:flipModel}).
    Numerical simulations are performed for different flipping frequencies $p=0.001, 0.003, 0.006$.
    (C) Cell proliferation model (\ref{eq:prolModel}).
    Numerical simulations are performed for different proliferation frequencies $q=0.001, 0.003, 0.006$.
    The expression patterns of the first $100$ cells are shown.
    Initial cell number $n=100$, time step $\Delta t = 0.01$, duration $t=1000$, and $(v, \beta, h, r) = (1, 100, 4, 40)$. Initial condition, $D_x(0)=D^0 + \kappa_x $ and $N_x(0)=N^0 + \kappa_x $, where $D^0$ and $N^0$ are the spatially homogeneous steady state values (Appendix A), and $\kappa_x$ is a random variable from the uniform distribution in $[-0.02,0.02]$).} 
    \label{fig:2}
\end{figure*}

We set the parameters $(v, \beta, h, r)$ such that linear analysis predicts the salt and pepper pattern when there is no cell rearrangement and we simulated the model (Fig.~\ref{fig:2}(A)). 
We then included cell rearrangement and found that the heterogeneity of the Delta-Notch pattern was decreased by cell rearrangement, and the homogeneous steady state became stable again for a sufficiently high level of cell rearrangement (Fig.~\ref{fig:2}(B)).
More precisely, when the flip frequency $p = 0.001$, the salt and pepper pattern was largely maintained. 
However, for increasing values of $p$, the amplitude of the pattern became smaller.
When $p$ was sufficiently large, the amplitude was almost $0$ for the whole region, and the system relaxed to the spatially homogeneous steady state (Fig.~\ref{fig:2}(B)).
Similar results were obtained with the cell proliferation model (Fig.~\ref{fig:2}(C)).
With increasing proliferation frequency $q$, the amplitude of the pattern became smaller and, finally, the system settled back to a homogeneous steady state.
These results are robust to $100$ different runs of numerical simulations for each parameter set.

To quantify the heterogeneity of the expression pattern, we introduce the heterogeneity function $H(t)$ as the variance of the Delta expression:
\begin{equation}\label{eq:Ht}
    H(t) = \frac{1}{n} \sum_{x=1}^n \left[ D_x(t)^2 - \langle D(t) \rangle^2 \right],
\end{equation}
where 
\begin{equation}\label{eq:avH}
    \langle D(t) \rangle = \frac{1}{n} \sum_{x=1}^n D_x(t).
\end{equation}
If the salt and pepper pattern is generated, then $H(t)$ is close to the squared value of the amplitude of the pattern. 
If Delta expression is spatially homogeneous at the steady state, then $H(t) = 0$.

\begin{figure}[htb]
    \includegraphics[width=8.6cm]{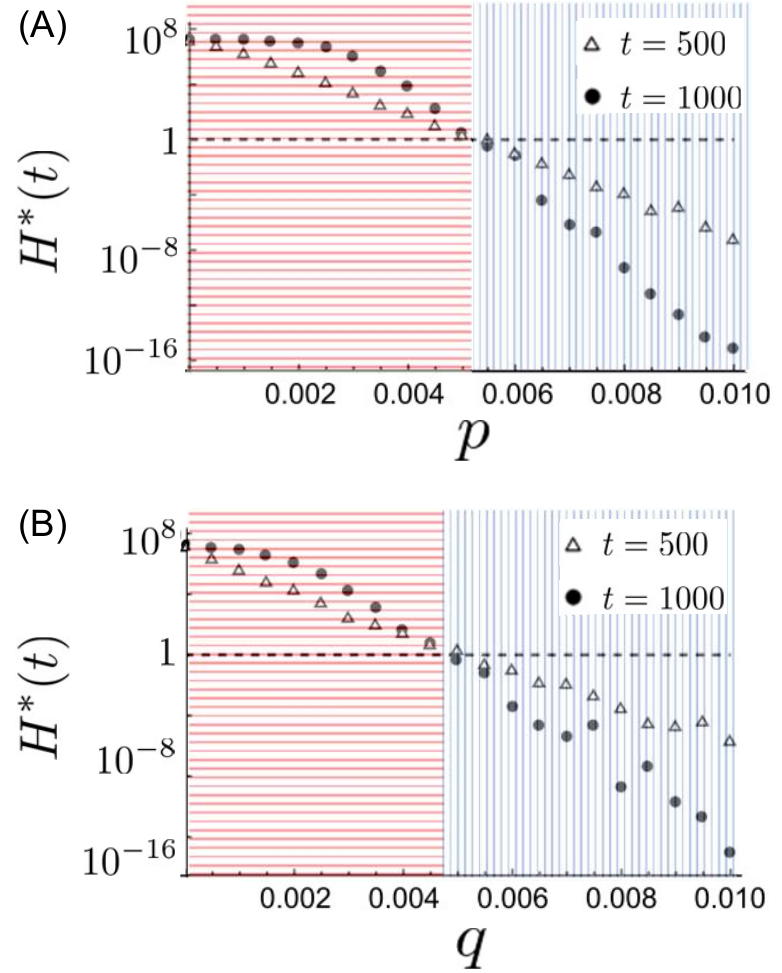}
    \caption{
    Log plots of the normalized heterogeneity of the pattern $H^*(t)$ against the frequencies of the cell rearrangement events for $t=500$ and $1000$.
    The black dashed line represents the plot of $H^*(t)=1$ and the circles and triangles represent $H^*(1000)$ and $H^*(500)$, respectively.
    (A) In the cell mixing model, $H^*(1000) > H^*(500) > H_0$ with $p \leq 0.05$ (red horizontal stripe region) and $H^*(1000) < H^*(500) < H_0$ with $p > 0.05$ (blue vertical stripe region).
    (B) In the cell proliferation model, similar inequalities hold, and the threshold value is $q = 0.045$.
    We calculated the heterogeneity at $21$ different frequencies of $p$ and $q$, which are taken in the range $0$ to $0.01$ at equal intervals of $0.0005$ in each model.
    The heterogeneity $H^*(t)$ shown in this figure was calculated by taking the average of $H(t)$ over $400$ different simulation runs, and then normalized by dividing by the average of $H_0 = H(10)$, for each $p$ and $q$.
    Other conditions are as in Fig.~\ref{fig:2}.
    Initial conditions are randomly determined from the same distribution as in Fig~\ref{fig:2} for each of the runs.
    }
    \label{fig:3}
\end{figure}

In both models, at the onset of the simulation, $H(t)$ decreases and then either increases or still decreases depending on the value of $p$ in the cell mixing model or the value of $q$ in the cell proliferation model (Fig.~S1).
This is because, at the onset, the initial random state is smoothened by the Delta-Notch dynamics.
As we are interested in pattern growth after a sufficient time has elapsed, we define $H_0 = H(10)$, and then define the normalized heterogeneity $H^*(t)$ as $H^*(t) = H(t)/H_0$, which is plotted in Fig.~\ref{fig:3}.

Figure~\ref{fig:3} shows that $H^*(t)$ switches between increasing and decreasing depending on the values of $p$ and $q$.
In the cell mixing model, it appears that $H^*(t)$ increases for $p \leq 0.05$ and decreases for $p > 0.05$ (Fig.~\ref{fig:3}(A) and Fig.~S1(A) \cite{SupMat}).
In the cell proliferation model, $H^*(t)$ increases for $q \leq 0.045$ and decreases for $q > 0.045$ (Fig.~\ref{fig:3}(B) and Fig.~S1(B) \cite{SupMat}).
These results suggest that there exist critical frequencies $p^*$ and $q^*$ for which the attenuation of the pattern by cell rearrangement and its formation by the Delta-Notch dynamics are balanced.

We numerically estimated the critical frequencies and the growth rate of the heterogeneity.
The critical frequencies $p^*$ and $q^*$ were estimated as the intersection points of the plot of $\ln{H^*(t)}$ as a function of $p$ and $q$ and the plot of $\ln{H^*(t)} = 0$ in Fig.~\ref{fig:3}.
The growth rate of the heterogeneity was also estimated by the slope of the line that was fitted to the plot of $\ln{H^*(t)}$ against $t$ in Fig.~S1 \cite{SupMat}.

\subsection{Analysis of the cell rearrangement models}
To quantify the effects of cell rearrangement, we analyzed the stability of the pattern dynamics and the critical frequencies $p^*$ and $q^*$.
The ``tug-of-war" of the cell rearrangement and the Delta-Notch dynamics was represented as the growth or attenuation of the heterogeneity $H(t)$.
Therefore, we focused on the effect of cell rearrangement on $H(t)$.

The heterogeneity $H(t)$ can also be calculated from the power spectrum of the Delta expression pattern.
The power spectrum $P_k$ of the Delta expression pattern are the squared absolute values of the Fourier coefficients $\delta_k$ of Delta expression (Appendix A), so $P_k$ can be calculated as:
\begin{equation}\label{eq:Pkt}
    P_k(t) = |\delta_k(t)|^2 = \left| \frac{1}{n} \sum_{x=1}^n D_x(t) e^{\frac{-i2\pi kx}{n}} \right|^2.
\end{equation}
Note that $k$ takes integer values from $0$ to $n-1$, and $n$ increases with time in the cell proliferation model.
From Parseval's theorem, 
\begin{equation}\label{eq:Parseval}
    \sum_{x=1}^n D_x(t)^2 = \sum_{k=0}^{n-1} P_k(t),
\end{equation}
and from (\ref{eq:Pkt}),
\begin{equation}\label{eq:P0t}
    \langle D_x(t) \rangle^2 = \left( \frac{1}{n} \sum_{x=1}^n D_x(t) \right)^2 = P_0(t).
\end{equation}
Therefore, by substituting (\ref{eq:Parseval}) and (\ref{eq:P0t}) into (\ref{eq:Ht}), $H(t)$ was calculated as below:
\begin{equation}
    H(t) = \frac{1}{n} \sum_{k=1}^{n-1} P_k(t).
\end{equation}

The critical frequencies $p^*$ and $q^*$ are independent of the definition of the heterogeneity $H(t)$.
If we adopted the variance of the Notch expression instead of the Delta expression, then the dispersion relation and the effect of the cell rearrangement events $A^j$ and $B^j$ are the same as for Delta expression, and we obtained the same $p^*$ and $q^*$ as before.
In addition, we can obtain the same $p^*$ and $q^*$ values if we defined the heterogeneity by the average of the squared values. 
For example, if we adopt $[\Sigma (D_x-D_{x+1})^2] /n$ as the heterogeneity, then we obtain the same $p^*$ and $q^*$ since this value is also calculated from the linear summation of the power spectrum (Fig. S2 \cite{SupMat}).
We now proceed to analyze the stability of the power spectrum $P_k(t)$ in the cell mixing and proliferation models.
\subsubsection{Cell mixing model}
First, we will transform the cell mixing model (\ref{eq:flipModel}) into the corresponding system of stochastic differential equations that represent the time evolution of the Fourier coefficients $\delta_k$.
To find the critical frequency $p^*$ and the onset of pattern formation, we assume that $H(t)$ is small since we set the initial condition to be a small perturbation about the homogeneous steady state, so the reaction terms $\boldsymbol{f}(\cdot), \boldsymbol{g}(\cdot)$ can be regarded as linear operators since $D_x \sim D^0$ and $N_x \sim N^0$.
Therefore, the effect of the Delta-Notch dynamics on the Fourier coefficients $\delta_k$ of $D_x$ is described by the diagonal matrix $\Lambda$ from the linear stability analysis (Appendix A) as below:
\begin{equation} \label{Diagonal}
    \Lambda = \text{Diag}(\lambda_0, \lambda_1, \cdots, \lambda_{n-1}),
\end{equation}
where
\begin{equation}\label{lambdak}
\lambda_k = \frac{-(a+d) + \sqrt{(a+d)^{2}-4(a d + 2b \alpha \cos{(2 \pi k/n})))}}{2}.
\end{equation}

The effect on the Fourier coefficients $\delta_k$ of a cell flip is given by the $n \times n$ matrix $C^j$:
\begin{equation} \label{eq:Cj}
    C^j = F A^j F^{-1},    
\end{equation}
where $F$ is the discrete Fourier transform matrix. 
The components of the matrices $F$ and $F^{-1}$ are given as below:
\begin{align}
    \left\{F\right\}_{l,m} = \frac{1}{\sqrt{n}} e^{-i2 \pi (l-1) (m-1)/n}, \\
    \left\{F^{-1}\right\}_{l,m} = \frac{1}{\sqrt{n}} e^{i2 \pi (l-1) (m-1)/n}. \label{eq:ff}
\end{align}
Therefore, the time evolution of the Fourier coefficients $\boldsymbol{\delta}$ can be described by:
\begin{equation}\label{eq:ddelta_flip}
    d \boldsymbol{\delta} = \Lambda \boldsymbol{\delta} dt + \sum_{j=1}^n (C^j-I) \boldsymbol{\delta} dL^{p,j}_t,
\end{equation}
where $\boldsymbol{\delta} = (\delta_0(t), \delta_1(t), \cdots, \delta_k(t), \cdots, \delta_{n-1}(t))^\mathsf{T}$.

Furthermore, we obtain the expected time evolution of the power spectrum by calculating the average of the effect of the cell flip on the power spectrum for $j$ (Appendix B) as below:
\begin{equation}\label{eq:flipdP}
    d \boldsymbol{P} = 2\mathrm{Re}[\Lambda] \boldsymbol{P} dt + W \boldsymbol{P} d L^{pn}_t.
\end{equation}
Here $\boldsymbol{P} = (P_0(t), P_1(t), \cdots, P_k(t), \cdots, P_{n-1}(t))^\mathsf{T}$, $L^{pn}_t$ is the Poisson process with intensity $p n$, and the components of the matrix $W$ are given as below:
\begin{equation}\displaystyle \label{eq:matW}
\left\{ W \right\}_{l,m} = 
\begin{cases}
 - \frac{8}{n} \sin^2{\frac{\pi (l-1)}{n}} + \left( \frac{4}{n} \sin^2{\frac{\pi (l-1)}{n}} \right)^2 & (l=m) \\
\left( \frac{4}{n} \sin{\frac{\pi (l-1)}{n}} \sin{\frac{\pi (m-1)}{n}} \right)^2 \ & (\rm{otherwise}) .
\end{cases}
\end{equation}

Both the average and variance of the Poisson process $L_t^{pn}$ are $pnt$, so those of $L_t^{pn}/n$ are $pt$ and $pt/n$, respectively.
Therefore, when $n$ is sufficiently large, $dL_t^{pn}/n$ can be approximated by $p dt$ and equation \eqref{eq:flipdP} is approximated by:
\begin{equation}\label{eq:2lpnw}
    \frac{d}{dt} \boldsymbol{P} \simeq Y_p \boldsymbol{P},
\end{equation}
where
\begin{equation}
    Y_p = 2\mathrm{Re}[\Lambda] + pn W .
\end{equation}
Therefore, by using the maximum eigenvalue and the corresponding eigenvector of the matrix $Y_p$, we can derive the expected pattern dynamics.

If $y$ is the maximum eigenvalue of ${Y}_p$ and $\boldsymbol{P^*}=(P^*_{0}, P^*_{1}, \cdots, P^*_{n-1})^\mathsf{T}$ is the corresponding eigenvector, then $\boldsymbol{P} \sim e^{yt} \boldsymbol{P^*}$ for values of $t$ in a range sufficiently large so that other eigenvectors no longer affect the power spectrum, but not so large for nonlinear effects to come into play.
The scaling law $H(t) \sim e^{yt}$ also holds since $H(t)$ is a linear summation of the power spectrum $P_k(t)$.
Therefore, the maximum eigenvalue $y$ corresponds to the growth rate of the heterogeneity $d \ln{H(t)}/dt$.
Figure~\ref{fig:4}(A) shows that the value of $y$ derived from equation \eqref{eq:2lpnw} agrees with the numerically estimated growth rate $d \ln{H(t)}/dt$, and Fig.~\ref{fig:4}(B) shows how the shape of the corresponding eigenvector $\boldsymbol{P^*}$ depends on $p$.
Note that the effect of the Delta-Notch interaction $2\mathrm{Re}[\Lambda]$ on $P_k$ is determined by the value of $2\pi k/n$, so we plot $P^*_k$ against $2\pi k/n$ in Fig.~\ref{fig:4}(B).

To obtain the critical frequency $p^*$, we used Newton's method to derive the value of $p$ such that the maximum eigenvalue of $Y_p$ is $0$.
Figure~\ref{fig:4}(C) shows that the values of $p^*$ obtained in this way are in very good agreement with the corresponding values estimated from the numerical simulations of (\ref{eq:flipModel}) for varying $\beta$.
The parameter $\beta$, which indicates the intensity of Delta suppression by Notch, broadens the region of positive $\lambda(\theta)$ and increases the maximum value of $\lambda(\theta)$ (Appendix A and Fig. S3).
The values of $y$ and $p^*$ obtained in Fig.~\ref{fig:4}(A,C) are almost identical for $n \geq 100$ (Fig.~S4 \cite{SupMat}).

Furthermore, we obtain the growth rate $d \ln{H(t)}/dt$ and the critical frequency $p^*$ as $n \to \infty$ as solutions of the integral equations (Supplemental text A \cite{SupMat}).
They are also in very good agreement with the numerically estimated values.

We can derive an approximation to the critical frequency $p^*$ from the linear stability analysis of the spatially uniform steady state in the deterministic system that is obtained by regarding the effect of cell mixing as a diffusion process:
\begin{align}\label{eq:diffusion}
    &\frac{d D_{x}}{d t}=v\left(\frac{1}{1+\beta N_{x}{ }^{h}}-D_{x}\right) + p(D_{x-1}+D_{x+1}-2 D_x) \nonumber\\
    &\frac{d N_{x}}{d t}=\frac{r\left(D_{x-1}+D_{x+1}\right)}{1+r\left(D_{x-1}+D_{x+1}\right)}-N_{x} + p(N_{x-1}+N_{x+1}-2 N_x).
\end{align}
System (\ref{eq:diffusion}) has the same spatially homogeneous steady state as in (\ref{eq:Collier}), so we can linearize the system as in Appendix A, and obtain the Jacobian matrix:
\begin{equation}
\tilde{M}_k =
\left(
    \begin{array}{cc}
    -a-4p \sin^2(\pi k/n) & -b \\
    2 \alpha \cos (2 \pi k / n) & -d - 4p \sin^2(\pi k/n)
    \end{array}
\right).
\end{equation}
The eigenvalue $\tilde{\lambda}_k$ with the larger real part, obtained from the matrix $\tilde{M}_k$, is:
\begin{equation}
    \tilde{\lambda}_k = \lambda_k - 4p \sin^2{\frac{\pi k}{n}},
\end{equation}
where $\lambda_k$ is given by \eqref{eq:lambdak2}, so the time evolution of the power spectrum can be approximated by:
\begin{equation}\label{eq:ignoreInt}
    \frac{d}{dt} P_k = \left(2\lambda_k - 8p \sin^2{\frac{\pi k}{n}} \right) P_k.
\end{equation}
This equation corresponds to the system that is obtained by ignoring the non-diagonal components of the matrix $Y_p$ in \eqref{eq:2lpnw}.
From (\ref{eq:ignoreInt}), the critical frequency $p^*$ is approximated as $p$ such that:
\begin{equation}\label{eq:maxlk}
    \underset{\theta \in [0,2\pi)}{\text{Max}} \left[ \lambda(\theta) -4 p \sin^2{\frac{\theta}{2}} \right] = 0.
\end{equation}
Figure~\ref{fig:4}(C) shows that the estimation in equation (\ref{eq:maxlk}) is a good approximation for $95< \beta < 120$.
If $\lambda(\theta)$ is positive only in the region that is very close to $\theta = \pi$, then we can obtain the simpler form of (\ref{eq:maxlk}):
\begin{equation}\label{eq:approx}
    p^* = \lambda_{\text{max}}/4.
\end{equation}
Here $\lambda_{\text{max}}$ is given by equation (\ref{eq:lambda}), and we used the approximation $\sin^2(\theta/2) \simeq 1$ in the region that is close to $\theta = \pi$.
Consistent with (\ref{eq:approx}), $\lambda_\text{max}$ was $0.02$ and the critical frequency $p^*$ was estimated around $0.005$ for the conditions used in Fig.~\ref{fig:3}.

\subsubsection{Cell proliferation model}
The cell proliferation model (\ref{eq:prolModel}) was also analytically transformed into the corresponding system of stochastic differential equations that represent the time evolution of $\delta_k$.
The effect of a cell proliferation event, which increases the cell number $n$ to $n+1$, on the Fourier coefficients $\boldsymbol{\delta}_n$, is given as below:
\begin{equation}\label{eq:chatj}
    \hat{C}^j = \hat{F} B_j {F}^{-1},
\end{equation}
where $\hat{F}$ is a square $(n+1) \times (n+1)$ matrix, ${F}^{-1}$ is the square $(n \times n)$ matrix defined in (\ref{eq:ff}), and $B_j$ is the $(n+1) \times n$ matrix given by \eqref{eq:matBj}. The matrix $\hat{F}$ is defined by:
\begin{equation}
    \left\{ \hat{F} \right\}_{l,m} = \frac{1}{\sqrt{n+1}} e^{-i2 \pi (l-1)(m-1)/(n+1)}.
\end{equation}
Therefore, the time evolution of $\boldsymbol{\delta}_n(t)$ is given as below:
\begin{equation}
    \begin{cases}
    \boldsymbol{\delta}_{n}(t+dt) = \mathrm{e}^{\Lambda dt} \boldsymbol{\delta}_n(t)  &\text {if } d L_{t}^{q, j}=0\\
    \boldsymbol{\delta}_{n+1}(t+d t)=\hat{C}^j \mathrm{e}^{\Lambda dt} \boldsymbol{\delta}_{n}(t) &\text {if } d L_{t}^{q, j}=1.
    \end{cases}
\end{equation}

By calculating the average of the effect of the cell proliferation event for $j$, the expected time evolution of the power spectrum $\boldsymbol{P}_n(t)$ is given (Appendix C) by:
\begin{equation} \label{eq:dPprol}
    \begin{cases}
        \boldsymbol{P}_{n}(t+dt) = \mathrm{e}^{2 \text{Re}[\Lambda] dt} \boldsymbol{P}_n(t) & \text{if } dL_t^{qn} =0 \\
        \boldsymbol{P}_{n+1}(t+dt) =  S \mathrm{e}^{2 \text{Re}[\Lambda] dt} \boldsymbol{P}_n(t) & \text{if } dL_t^{qn} =1,
    \end{cases}
\end{equation}
where the components of the matrix $S$ are given by:
\begin{equation}\displaystyle \label{eq:matS}
    \left\{ S \right\}_{l,m} = 
    \begin{cases}
        (n+1)/n & (\text{if } l=m=1) \\
        \frac{1}{n(n+1)} \frac{\sin^2{\frac{\pi m}{n}}}{\sin^2 (\frac{\pi l}{n+1}- \frac{\pi m}{n})} & (\text{otherwise}).
    \end{cases}
\end{equation}

Since the matrix $S$ is non-square, the stability of the homogeneous steady state cannot be determined as in the cell mixing model.
Hence we approximate the matrix $S$ by a square matrix, as below.

The power spectrum $\boldsymbol{P}_n$ is represented by the superposition of the cosine waves from the symmetry $P_k=P_{n-k}$.
By assuming that the shortest wavelength component of $S\boldsymbol{P}_n$ is negligible, the matrix $S$ is approximated by the square matrix $\Sigma$ (given below) and the equation \eqref{eq:dPprol} is approximated by (Appendix C):
\begin{equation}\label{eq:sigma}
    d\boldsymbol{P}_n \simeq 2 \text{Re}[\Lambda] \boldsymbol{P}_n dt + (\Sigma-I) \boldsymbol{P}_n dL_t^{qn},
\end{equation}
where $I$ is the identity matrix.
When $n$ is even, the components of the matrix $\Sigma$ are given by:
\begin{align}\displaystyle \label{eq:matSigma}
    \left\{ \Sigma \right\}_{l,m} = 
    &\frac{2}{n} \sum_{k=2}^{{n}/{2}} \cos \frac{2 \pi(m-1)(k-1)}{n}\left[\frac{k-1}{n+1} \cos \frac{2 \pi(l-1)(k-2)}{n}+\left(1-\frac{k-1}{n+1}\right) \cos \frac{2 \pi(l-1)(k-1)}{n}\right] \nonumber\\
    &+\frac{1}{n}\left(1+(-1)^{m+k-2}\left(1-\frac{n}{n+1} \sin ^{2} \frac{\pi(m-1)}{n}\right)\right),
\end{align}
and when $n$ is odd:
\begin{align}\displaystyle \label{eq:matSigmaOdd}
    \left\{ \Sigma \right\}_{l,m} = 
    &\frac{2}{n} \sum_{k=2}^{{(n+1)}/{2}} \cos \frac{2 \pi(m-1)(k-1)}{n}\left[\frac{k-1}{n+1} \cos \frac{2 \pi(l-1)(k-2)}{n}+\left(1-\frac{k-1}{n+1}\right) \cos \frac{2 \pi(l-1)(k-1)}{n}\right] \nonumber \\
    &+\frac{1}{n} .
\end{align}

As in the cell mixing model, assuming $n$ is sufficiently large, $L_t^{qn}/n$ is approximated by $qt$, so the time evolution of $\boldsymbol{P}_n$ in \eqref{eq:sigma} is approximated by:
\begin{equation}\label{eq:matJ}
    \frac{d}{dt} \boldsymbol{P}_n \simeq J_q \boldsymbol{P}_n,
\end{equation}
where
\begin{equation}
    J_q = 2 \text{Re}[\Lambda] + qn (\Sigma-I).
\end{equation}

Therefore, by using the maximum eigenvalue and the corresponding eigenvector of the matrix $J_q$, we can approximately derive the expected pattern dynamics.

\begin{figure*}[htb]
    \includegraphics[width=17.8cm]{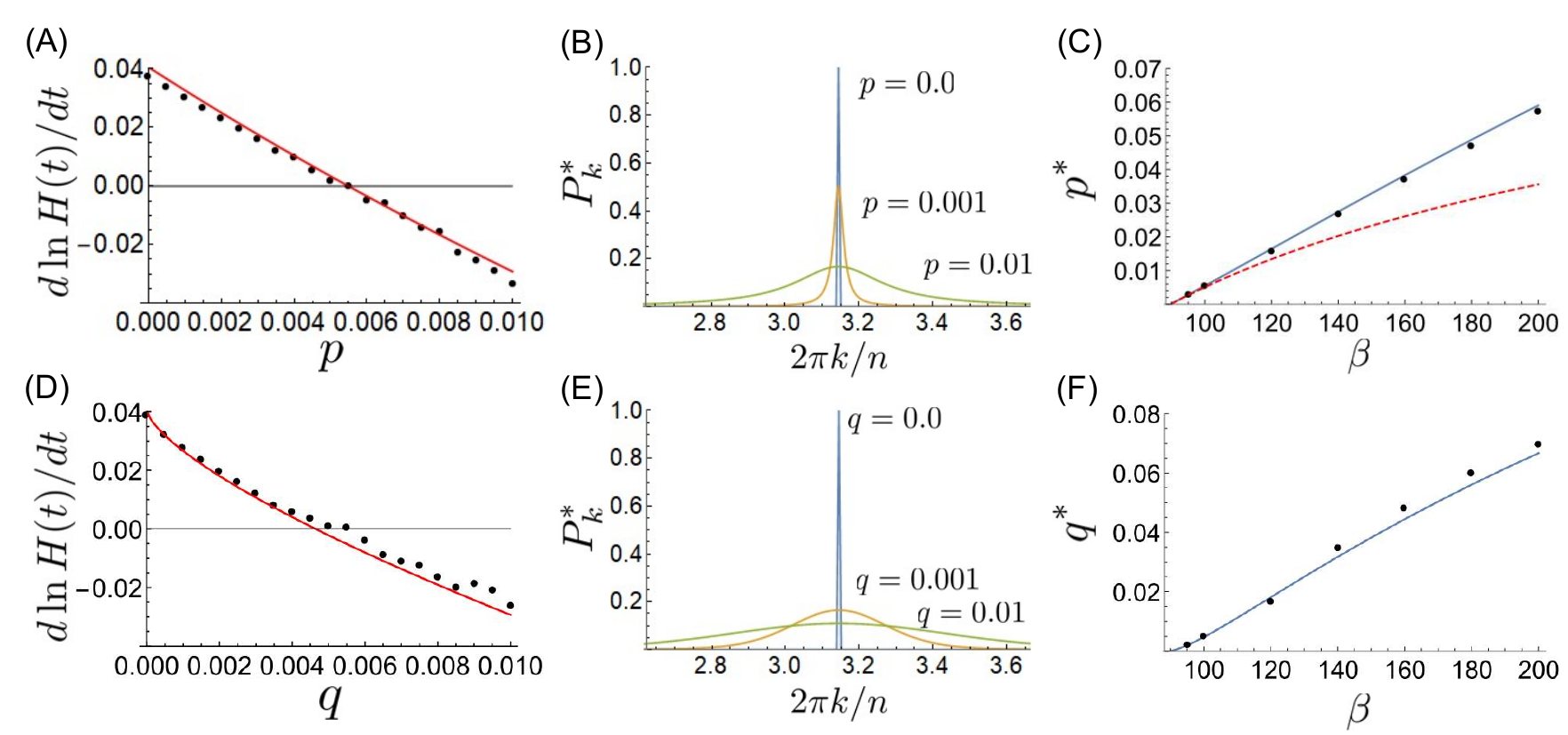}
    \caption{
        Comparison between the analytical and numerical results.
        (A) The red line and the black dots represent the maximum eigenvalue $y$ of the matrix $Y_p$ in \eqref{eq:2lpnw} and the growth rate $d \ln{H(t)}/dt$ estimated from Fig.~S1(A) \cite{SupMat}, respectively.
        (B) Normalized corresponding eigenvector $\boldsymbol{P^*}$ with the maximum eigenvalue of the matrix ${Y}_p$ with $n=1000$.
        (C) Critical frequencies $p^*$ plotted against the parameter $\beta$ in the Collier model (\ref{eq:Collier}).
        The blue solid line represents the values of $p$ such that the maximum eigenvalue of $Y_p$ in \eqref{eq:2lpnw} is $0$, the red dashed line represents $p^*$ derived from (\ref{eq:maxlk}) and the black dots represent the values of $p^*$ that were estimated from Fig.~\ref{fig:3}(A).
        (D) The red line and the black dots represent the maximum eigenvalue of the matrix $J_q$ in \eqref{eq:matJ} and the growth rate $d \ln{H(t)}/dt$ estimated from Fig.~S1(B) \cite{SupMat}, respectively. 
        (E) Normalized corresponding eigenvector $\boldsymbol{P^*}_n$ with the maximum eigenvalue of the matrix ${J}_q$ with $n=1000$.
        (F) Critical frequencies $q^*$ plotted against the parameter $\beta$.
        The blue line represents the value of $q$ such that the maximum eigenvalue of $J_q$ in \eqref{eq:matJ} is $0$, and the black dots represent the values of $q^*$ that were estimated from Fig.~\ref{fig:3}(B), respectively.
        The numerically estimated growth rate $d \ln{H(t)}/dt$ (black dots in (A) and (D)) were calculated from the slope of the lines that were fitted to the plot of $\ln{H(t)}$ against $t$ (Fig. S1 \cite{SupMat}).
        The numerically estimated critical frequencies (black dots in (C) and (F)) were estimated as the intersection points of the plot of $\ln{H^*(t)}$ as a function of $p$ and $q$ and the plot of $\ln{H^*(t)} = 0$ in Fig.~\ref{fig:3}, respectively.
    }
    \label{fig:4}
\end{figure*}

Figure~\ref{fig:4}(D) shows that the maximum eigenvalue of the matrix $J_q$ is in very good agreement with the numerically estimated growth rate $d \ln{H(t)}/dt$, and Fig.~\ref{fig:4}(E) shows how the shape of the corresponding eigenvector $\boldsymbol{P^*}_n$ depends on $q$.

To obtain the critical frequency $q^*$, we used Newton's method to derive the value of $q$ such that the maximum eigenvalue of $J_q$ is $0$.
Figure~\ref{fig:4}(F) shows that the values of $q^*$ obtained in this way are in very good agreement with the numerically estimated $q^*$.
The values obtained in Fig.~\ref{fig:4}(D,F) are almost identical for $n \geq 100$ (Fig.~S4 \cite{SupMat}).

In summary, we have shown that the pattern dynamics of the cell mixing model and the cell proliferation model, analyzed by solving the maximum eigenvalue problem of the matrices ${Y}_p$ and ${J}_q$, respectively, yield results on the critical frequencies that agree well with our numerical simulations of the full model.

\subsection{Experimental results}\label{sec:exp}
\begin{figure*}[htb]
    \includegraphics[width=17.8cm]{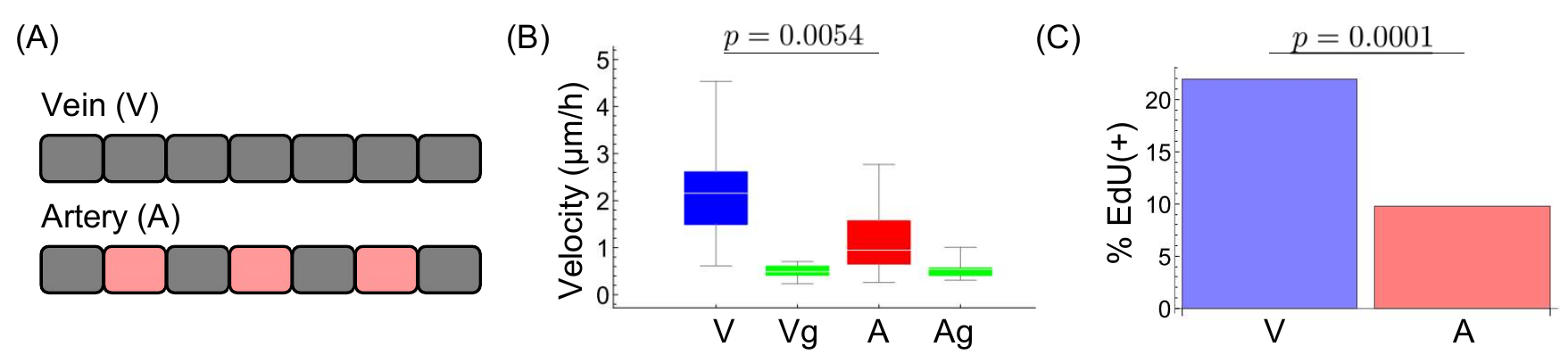}
    \caption{
    Experimental assay of cell movement and proliferation.
    (A) Schematic of Dll4 expression of endothelial cells in the retinal vasculature.
    The grey rectangles represent cells having low Dll4 expression and the pink rectangles represent cells having high Dll4 expression.
    Dll4 expressing cells are rarely observed in veins, while they are alternately aligned in arteries \cite{claxton_periodic_2004, hofmann_notch_2006, herman_novel_2018}.
    (B) Box chart of the averaged velocities of the endothelial cells.
    Endothelial cells move at an average speed of $2.19$ {\textmu}m/h in veins ($n=24$, denoted by ``V''), $1.08$ {\textmu}m/h in arteries ($n=13$, denoted by ``A'') and ganglion cells move at $0.49$ {\textmu}m/h near veins ($n=10$, denoted by ``Vg'') and $0.56$ {\textmu}m/h near arteries ($n=10$, denoted by ``Ag'').
    P-value between endothelial cells in the veins and in the arteries is $0.0054$ (Student t-test).
    (C) Bar chart of the fraction of the EdU(+) endothelial cells.
    $21.9 \%$ (93/424) of venous endothelial cells and $9.8 \%$ (20/204) of arterial endothelial cells are positive for EdU.
    P-value is $0.0001$ (Fisher's exact test).
    }
    \label{fig:5}
\end{figure*}

An example of Delta-Notch pattern formation on a one-dimensional line is the expression pattern of Delta-like ligand 4 (Dll4) mRNA in retinal endothelial cells, as reported in \cite{claxton_periodic_2004, hofmann_notch_2006, herman_novel_2018}.
Delta-like ligand 4 (Dll4) is a Notch ligand that is expressed mainly in the blood vessels, specifies the tip cells of growing vessels, and promotes arterialization \cite{herman_novel_2018}.
In the retinal arteries, Dll4 positive and negative cells alternately align, though in veins the expression level is low and almost homogeneous (Fig.~\ref{fig:5}(A)).
In addition, cell mixing and proliferation of endothelial cells in the developing vasculature are reported in \cite{arima_angiogenic_2011, pontes-quero_high_2019, luo_arterialization_2020}.
Therefore, the developing retinal vasculature is a good example on which to test our model \textit{in vivo}.

We examined the motility of the endothelial cells in veins and arteries by using an \textit{ex vivo} assay, corresponding to the cell mixing model (\ref{eq:flipModel}).
We enucleated the eyes from the P5 mouse and isolated the retina.
We cultured the retinal explants and tracked the movement of endothelial cells in the developing retinal vasculature for $12$ h (Fig.~S5(A--D) \cite{SupMat}).
In veins, endothelial cells actively migrated and exchanged their relative positions (Fig. S5(C, E) \cite{SupMat}), though in the arteries the motility of the endothelial cells was lower and they rarely exchanged positions (Fig.~S5(D) \cite{SupMat}).
Cell displacements at $12$ h were quantified, and the velocity of endothelial cells, which was calculated by dividing displacement by the observation time, was found to be $2.19$ {\textmu}m/h in veins and $1.08$ {\textmu}m/h in arteries (Fig.~\ref{fig:5}(B)).
These velocities are significantly different from each other ($\text{p-value}=0.0054$).
Note that the velocities of the ganglion cells were much smaller than those of the endothelial cells (Fig.~\ref{fig:5}(B)), supporting the hypothesis that the motion of the endothelial cells is active and did not arise passively from the deformation of the surrounding tissues.

Next, we examined the proliferation rate of endothelial cells in veins and arteries by \textit{in vivo} assay, corresponding to the cell proliferation model (\ref{eq:prolModel}).
Proliferation was assessed by EdU incorporation in $2$ h in the postnatal mouse retinal vasculature \textit{in vivo} (Fig.~\ref{fig:5}(C) and Fig.~S6 \cite{SupMat}).
We intraperitoneally injected EdU $2$ h before sacrifice and detected EdU incorporating cells.
Since EdU was selectively incorporated by cells in the S phase, we could use this to identify proliferating cells.
The fraction of EdU positive cells among endothelial cells, distinguished by the expression of ERG1, was $21.9 \%$ in veins and $9.8 \%$ in arteries.
These percentages of EdU positive cells are significantly different from each other ($\text{p-value} =0.0001$).

Our theoretical results showed that the critical frequency $q^*$ in the cell proliferation model is of a similar order of magnitude as that of the growth rate of pattern formation $\lambda_\text{max}$ (Fig.~\ref{fig:4}(F)).
The proliferation rate of endothelial cells in veins is estimated to be on the order of 1/10 $\text{h}^{-1}$ since $21.9 \%$ of the cells entered the S phase in 2 hours in Fig.~\ref{fig:5}.
The characteristic time scale of Delta-Notch pattern formation, which corresponds to $1/\lambda_\text{max}$, is estimated to be about 10 $\text{h}$ according to reports \cite{hasan_endothelial_2017}.
This result implied that the critical frequency $q^*$ could be between the observed proliferation rate of arterial and venous endothelial cells, and the differences in the proliferation rate could account for the difference in the Dll4 expression pattern.

We conclude that both motility and proliferation rate of endothelial cells are higher in veins, in which the Dll4 expression pattern is spatially homogeneous, than in arteries, which shows the salt and pepper pattern.
These results are quantitatively consistent with our model and suggest the possibility that cell rearrangement events may affect the Delta-Notch pattern \textit{in vivo}.

\section{Discussion}
To our knowledge, this paper is the first to provide a framework to analytically evaluate the effect on Delta-Notch pattern formation of cell rearrangement arising from migration or proliferation in a one-dimensional line of cells.
We model cell rearrangement events as occurring intermittently and randomly in a discrete spatial linear structure.
We modeled the intermittency of cell rearrangement events by a jump process and analyzed the model while maintaining the discreteness of the spatial structure by considering the time evolution of the power spectrum.
In our framework, the stochastic and intermittent effects of cell rearrangement were approximated by the deterministic effects on the power spectrum.
Accordingly, the instabilities of the pattern dynamics were analyzed by solving the maximum eigenvalue problem of the resultant systems \eqref{eq:2lpnw} and \eqref{eq:matJ}.

Although both cell mixing and proliferation decrease the amplitude of the Delta-Notch pattern in our model (Fig.~\ref{fig:2}), their detailed effects on the power spectrum are qualitatively different.
In frequency space, both effects are regarded as the redistribution of the power spectrum as $W$ and $S$ in equations (\ref{eq:matW}) and (\ref{eq:matS}), respectively.
Cell mixing coarsens the power spectrum to be uniformly distributed, while cell proliferation shifts the distribution of the power spectrum to the long-wavelength region.
These effects correspond, in the absence of cell-cell interaction, to the scrambling of existing patterns due to cell mixing, and elongation of an existing pattern due to cell proliferation.

In this study, our results suggest that the Delta expression pattern of cells is transformed from one in which cells with high Delta expression and those with low Delta expression are alternately aligned, to a pattern in which all cells express moderate levels of Delta and are spatially homogeneously aligned, depending on the frequency of cell rearrangement events (Fig. \ref{fig:2}).
Conversely, Pontes-quero \textit{et al}. \cite{pontes-quero_high_2019} proposed that Notch signaling regulates endothelial cell proliferation in a bell-shaped dose-response manner, which means that sufficiently high or low Notch signaling inhibits cell proliferation, and moderate Notch signaling promotes it. 
These facts imply that the proliferation rate of cells is inhibited in the salt and pepper pattern and promoted in the spatially homogeneous pattern, and there is a feedback loop between cell proliferation and Delta-Notch pattern formation.
In the salt and pepper pattern, cell proliferation is inhibited, and a low proliferation rate contributes to salt and pepper pattern formation.
Conversely, in the homogeneous pattern case (with moderate levels of Notch expression), cell proliferation is promoted, and a high proliferation rate contributes to the homogeneous pattern.
In addition, Luo \textit{et al}. reported that inhibition of cell proliferation by Dll4-Notch1 signaling is essential for arterialization from the rudimentary plexus \cite{luo_arterialization_2020}. 
The feedback loop comprising the Delta-Notch system and cell proliferation may contribute to the establishment of vessel differentiation by constituting a bistable system consisting of the low proliferative salt and pepper pattern, and the high proliferative homogeneous pattern, respectively, arteries and veins.

For Delta-Notch pattern formation, additional mechanisms that are not included in the Collier model (\ref{eq:Collier}) have recently been reported.
For example, Sprinzak \textit{et al}. showed that the Delta ligand inhibits Notch in single cells (cis-interaction) \cite{sprinzak_cis-interactions_2010}, and Nandagopal \textit{et al}. showed that the Delta ligand could also activate Notch depending on the Delta concentration (cis-activation) \cite{nandagopal_cis-activation_2019}.
However, a full model of the Delta-Notch system is still not established.
In this work, we used the Collier model (\ref{eq:Collier}) because this model is a simple model that can capture many features of Delta-Notch pattern formation.
However, importantly, we used only the dispersion relation of the Delta-Notch system to derive the critical frequencies in equations (\ref{eq:2lpnw}) and (\ref{eq:matJ}).
Therefore, our framework can also be applicable to other pattern formation mechanisms, including the modified Delta-Notch model, to investigate the robustness and feasibility of the pattern dynamics.

\section*{Acknowledgements}
The authors thank Professor Koichi Nishiyama (Kumamoto University), Professor Shin-Ichiro Ei (Hokkaido University), Dr. Yoshitaro Tanaka (Future University Hakodate), and Dr. Kei Sugihara (Kyushu University) for helpful discussions. This work is financially supported by JST CREST (Grant Number JPMJCR14W4) and TOBITATE! Study Abroad Initiative and Young Ambassador Program.

\onecolumngrid
\appendix
\section{Dispersion-relation of the Collier model}
To derive the necessary and sufficient conditions for pattern formation, we performed a linear stability analysis of the Collier model (\ref{eq:Collier}).

The homogeneous steady state $(D^0, N^0)$ in the Collier model (\ref{eq:Collier}) with periodic boundary conditions is given by:
\begin{align}
    {D}^0 = \frac{1}{1+ \beta {({N}^0})^h}\label{eq:dcline} \\
    {N}^0 = \frac{2 r D^0 }{1+ 2 r D^0} . \label{eq:ncline}
\end{align}

By setting $D_x = D^0 + d_x, N_x = N^0 + n_x$, where $|d_x| \ll 1, |n_x| \ll 1$, the Collier model (\ref{eq:Collier}) can be linearized to obtain:
\begin{align}\label{eq:linearModel}
\frac{d}{dt} d_x &= -a d_x - b n_x \nonumber \\ 
\frac{d}{dt} n_x &= -d n_x + \alpha (d_{x-1}+d_{x+1}),
\end{align}
where $a = v, \ b = (\beta h v (N^{0})^{h-1})/\left(1+\beta (N^{0})^h \right)^{2}, \ d = 1, \ \alpha = {r}/\left(1+2 r D^{0}\right)^{2}$.

To examine the stability of the homogeneous steady state in the Collier model (\ref{eq:Collier}), we consider a discrete Fourier transformation of $d_x, n_x$ as below:
\begin{align}\displaystyle \label{eq:DFT}
\delta_k (t)&= \frac{1}{\sqrt{n}} \sum_{x=1}^{n} d_x(t) \mathrm{e}^{i2\pi kx/n} \nonumber \\
\nu_k (t)&= \frac{1}{\sqrt{n}} \sum_{x=1}^{n} n_x(t) \mathrm{e}^{i2\pi kx/n},
\end{align}
where,
\begin{align}\displaystyle \label{eq:iDFT}
    d_x (t)&= \frac{1}{\sqrt{n}} \sum_{k=0}^{n-1} \delta_k(t) \mathrm{e}^{- i2\pi kx/n} \nonumber \\
    n_x (t)&= \frac{1}{\sqrt{n}} \sum_{k=0}^{n-1} \nu_k(t) \mathrm{e}^{- i2\pi kx/n}. 
\end{align}
Here, $k$ is the wavenumber and takes integer values from $0$ to $n-1$, while $\delta_k (t)$ and $\nu_k (t)$ are the Fourier coefficients that take complex values.

Substituting (\ref{eq:iDFT}) into (\ref{eq:linearModel}), we obtain a system of ordinary differential equations for the coefficients $\delta_k$ and $\nu_k$ as below:
\begin{equation}\label{eq:Matrixform}\displaystyle
\frac{d}{dt} 
\left(
    \begin{array}{c}
      \delta_k (t) \\
      \nu_k (t)
    \end{array}
    \right)
= 
M_k
\left(
    \begin{array}{c}
      \delta_k (t) \\
      \nu_k (t)
    \end{array}
    \right),
    \end{equation}
where
\begin{equation} \displaystyle \label{eq:Jacobian}
M_k =
\left(
    \begin{array}{cc}
      -a & -b \\
      2\alpha \cos{(2\pi k/n)} & -d 
    \end{array}
  \right).
\end{equation}

Setting
\begin{equation}
\left(
    \begin{array}{c}
      \delta_k (t) \\
      \nu_k (t)
    \end{array}
    \right)
=  
\left(
    \begin{array}{c}
      \delta_k (0) \\
      \nu_k (0)
    \end{array}
    \right)
    \mathrm{e}^{\lambda_k t},
\end{equation}
we find that $\lambda_k$ is an eigenvalue of $M_k$, and the solution is dominated by the larger eigenvalue of the Jacobian matrix $M_k$ (if both eigenvalues are real).
Therefore, whether the components $\delta_k, \nu_k$ grow or decay is determined by the sign of $\lambda_k$, where:
\begin{equation}\label{eq:lambdak2}
\lambda_k = \frac{-(a+d) + \sqrt{(a+d)^{2}-4(a d + 2b \alpha \cos{(2 \pi k/n})))}}{2}.
\end{equation}
Note that if $\lambda_k$ is complex, then the real part of $\lambda_k$ is negative and so the perturbation decays with time.
In the Collier model (\ref{eq:Collier}), $\lambda_k$ in (\ref{eq:lambdak2}) takes its largest value at $k=n/2$, which corresponds to the salt and pepper pattern, and the necessary and sufficient condition for pattern formation is obtained as below:
\begin{equation}\label{eq:lambda2}
\lambda_\text{max} = \frac{-(a+d) + \sqrt{(a+d)^{2}-4(a d - 2b \alpha))}}{2} > 0.
\end{equation}

From (\ref{eq:dcline}) and (\ref{eq:ncline}), we have that,
\begin{equation}
    \beta (N^0)^{h+1} = - (2r+1) N^0 + 2r.
\end{equation}
Thus,
\begin{align}
    b &= \frac{hv (2r-(2r+1)N^0}{4r^2 (1-N^0)^2} \\
    \alpha &= (1-N^0)^2 r,
\end{align}
and
\begin{equation}
    2 b \alpha = h v \left(1- N^0 - \frac{N^0}{2r} \right).
\end{equation}
Since $ad = v$ and $0 < N^0 < 1$ from (\ref{eq:ncline}), $ad > 2b\alpha$ if $h \leq 1$, so that the inequality (\ref{eq:lambda2}) does not hold.
Hence a necessary condition for (\ref{eq:lambda2}) to hold is $h>1$.

\section{Derivation of the time evolution equation for the power spectrum (\ref{eq:flipdP})}
From equation (\ref{eq:ddelta_flip}), the value of $\delta_k(t+dt)$ is given by:
\begin{equation}\label{eq:B1}
    \delta_k(t+dt) = \delta_k(t) + \lambda_k \delta_k(t) dt + \sum_{j=1}^{n} \sum_{l=0}^{n-1} \left\{C^j - I \right\}_{k+1,l+1} \delta_l(t) dL^{p,j}_t.
\end{equation}
The value of the power spectrum $P_k(t+dt) = |\delta_k(t+dt)|^2$ is obtained by multiplying ${\delta}_k(t+dt)$ in \eqref{eq:B1} by its complex conjugate $\overline{\delta}_k(t+dt)$ as below:
\begin{align} \label{eq:flip_org}
    \left|\delta_{k}(t+dt)\right|^{2} =& \left|\delta_{k}(t)\right|^{2} + \lambda_{k}\left|\delta_{k}(t)\right|^{2} dt + \overline{\lambda}_{k} \left|\delta_{k}(t)\right|^{2} dt \nonumber \\
    &+ \sum_{j=1}^n 
    \left[ 
    \left( \overline{\delta}_{k}(t) \sum_{l=0}^{n-1} \left\{C^j-I \right\}_{k+1,l+1} \delta_{l}(t)+\delta_{k}(t) \sum_{l=0}^{n-1} \left\{ \overline{C^j}-I \right\}_{k+1,l+1} \overline{\delta}_{l}(t) 
    \right) \right. \nonumber \\
    &\left. \quad +\left(\sum_{l=0}^{n-1} \left\{C^j-I \right\}_{k+1,l+1} \delta_{l}(t)\right)
    \left(
    \sum_{l=0}^{n-1} \left\{ \overline{C^j}-I \right\}_{k+1,l+1} \overline{\delta}_{l}(t)
    \right) 
     \right] d L^{p,j}_{t} \nonumber \\
    &+ O(dL^{p,j} \ dt) + O(dt^2).
\end{align}
Here we used the result:
\begin{equation}
    (dL^{p,j}_t)(dL^{p,\xi}_t) =
    \begin{cases}
        0 & \text{if } j \neq \xi \\
        dL^{p,j}_t & \text{if } j = \xi.
    \end{cases}
\end{equation}
By denoting $a_{k}^{j}=\sum_{l=0}^{n-1}\left\{C^{j}\right\}_{k+1, l+1} \delta_{l}$, we obtain:
\begin{equation} \label{eq:ajk}
    \sum_{l=0}^{n-1}\left\{C^{j}-I\right\}_{k+1, l+1} \delta_{l}(t)=a_{k}^{j}(t)-\delta_{k}(t).
\end{equation}
Substituting (\ref{eq:ajk}) into (\ref{eq:flip_org}), we obtain:
\begin{align} \label{eq:flipAll}
    \left|\delta_{k}(t+dt)\right|^{2} =& \left|\delta_{k}(t)\right|^{2} + 2 \text{Re}[\lambda_k] \left|\delta_{k}(t)\right|^{2} dt + \sum_{j=1}^{n} \left[ |a^j_k(t)|^2 - |\delta_k(t)|^2 \right] dL^{p,j}_{t} \nonumber \\
    =& \left|\delta_{k}(t)\right|^{2} + 2 \text{Re}[\lambda_k] \left|\delta_{k}(t)\right|^{2} dt
    +\sum_{j=1}^n 
    \left[ \left|\sum_{l=0}^{n-1} \left\{C^j \right\}_{k+1,l+1}{\delta}_{l}(t) \right|^2 - |{\delta}_{k}(t)|^2
     \right] d L^{p,j}_{t}.
\end{align}

The third term on the right-hand side of \eqref{eq:flipAll} is the effect of the cell flip on the power spectrum for the flip position $j$.
Based on the symmetry of the cell position $j$ in the system (\ref{eq:flipModel}), we assume that the third term on the right-hand side of \eqref{eq:flipAll} is approximated by replacing the effect of each flip event with $\mathcal{W}_k$, which is the averaged effect for the flip position $j$ as below:
\begin{equation} \label{eq:flipReplace}
    \sum_{j=1}^n 
    \left[ 
        \left|\sum_{l=0}^{n-1} \left\{C^j \right\}_{k+1,l+1}{\delta}_{l}(t) \right|^2 - |{\delta}_{k}(t)|^2
    \right] d L^{p,j}_{t}
    \simeq  \sum_{j=1}^n \mathcal{W}_k  d L^{p,j}_t 
    =       \mathcal{W}_k d L^{pn}_t,
\end{equation}
where,
\begin{align} \label{eq:W_k}
    \mathcal{W}_k 
    &=
        \frac{1}{n} \sum_{j=1}^n 
        \left[
            \left| \sum_{l=0}^{n-1} \left\{C^j \right\}_{k+1,l+1}{\delta}_{l}(t) \right|^2 - |{\delta}_{k}(t)|^2
        \right] \nonumber \\
    &=
    \frac{1}{n} \sum_{j=1}^n 
    \left[
        \left| \sum_{l=0}^{n-1} \left\{C^j \right\}_{k+1,l+1}{\delta}_{l}(t) \right|^2
    \right] 
    - |{\delta}_{k}(t)|^2
\end{align}
Here, we used $\sum_{j=1}^n dL_t^{p,j} = dL_t^{pn}$ and note that
\begin{align}\label{eq:aveapprox}
    \frac{1}{n} \sum_{j=1}^n 
    \left[ 
        \left| \sum_{l=0}^{n-1} \left\{C^j \right\}_{k+1,l+1}{\delta}_{l}(t) \right|^2 
    \right]
    = \frac{1}{n} \sum_{m=0}^{n-1} \sum_{l=0}^{n-1}
    \left[
        \delta_{l} \overline{\delta_{m}} 
        \left( 
            \sum_{j=1}^{n} \left\{C^{j}\right\}_{k+1, l+1} 
            \overline{\left\{C^{j}\right\}}_{k+1, m+1} 
        \right)
    \right].
\end{align}

From (\ref{eq:Cj}), the components of the matrix $C^j$ are given as below:
\begin{equation}
    \left\{C^j \right\}_{k,l} = 
    \begin{cases}
        -\frac{4}{n}  \sin{\frac{\pi(l-1)}{n}} \sin{\frac{\pi(k-1)}{n}}e^{\frac{i\pi(2j-1)(k-l)}{n}} & \text{if } k \neq l \\
        1- \frac{4}{n}  \sin^2{\frac{\pi(k-1)}{n}} & \text{if } k = l,
    \end{cases}
\end{equation}
so,
\begin{equation}\label{eq:cjcj}
    \sum_{j=1}^{n} \left\{C^{j}\right\}_{k+1, l+1} 
            \overline{\left\{C^{j}\right\}}_{k+1, m+1}
    =
    \begin{cases}
        0 & \text{if } l \neq m \\
        \frac{16}{n} \sin^2{\frac{\pi k}{n}} \sin^2{\frac{\pi l}{n}} & \text{if } l=m \text{ and } k \neq m \\
        n \left( 1-\frac{4}{n} \sin^2{\frac{\pi k}{n}} \right)^2 & \text{if } k=l=m.
    \end{cases}
\end{equation}
Here we used
\begin{equation}
    \sum_{j=1}^n e^{i\pi (2j-1)(l-m)/n} =
    \begin{cases}
        0 & \text{if } l \neq m \\
        n & \text{if } l = m.
    \end{cases}
\end{equation}
Therefore, from (\ref{eq:aveapprox}) and (\ref{eq:cjcj}), we obtain:
\begin{align} \label{eq:AverageFlip}
    &\frac{1}{n} \sum_{j=1}^n 
    \left[ 
        \left| \sum_{l=0}^{n-1} \left\{C^j \right\}_{k+1,l+1} {\delta}_{l}(t) \right|^2 
    \right] \nonumber \\
    & = \frac{1}{n} \sum_{l=0 \atop l \neq k}^{n-1} \left[ \left|\delta_l(t)\right|^2 \frac{16}{n} \sin^2{\frac{\pi k}{n}} \sin^2{\frac{\pi l}{n}} \right] + \left| \delta_k(t) \right|^2 \left( 1- \frac{8}{n} \sin^2{\frac{\pi k}{n}} + \frac{16}{n^2} \sin^4{\frac{\pi k}{n}} \right) \nonumber \\
    &= \sum_{l=0}^{n-1} \left[ \left( \frac{4}{n} \sin{\frac{\pi k}{n}} \sin{\frac{\pi l}{n}}\right)^2
    |\delta_l(t)|^2 \right] + \left( 1-\frac{8}{n} \sin^2{\frac{\pi k}{n}} \right) |\delta_k(t)|^2.
\end{align}
By replacing the third term on the right-hand side of (\ref{eq:flipAll}) by the averaged effect (\ref{eq:flipReplace}) and substituting \eqref{eq:W_k} and \eqref{eq:AverageFlip}, we obtain:
\begin{align}
    \left|\delta_{k}(t+dt)\right|^{2} 
    \simeq & \left|\delta_{k}(t)\right|^{2} + 2 \text{Re}[\lambda_k] \left|\delta_{k}(t)\right|^{2} dt
    + \frac{1}{n} \sum_{j=1}^n 
    \left[ \left|\sum_{l=0}^{n-1} \left\{C^j \right\}_{k+1,l+1}{\delta}_{l}(t) \right|^2 - |{\delta}_{k}(t)|^2
     \right] d L^{pn}_{t} \nonumber \\
     =& \left|\delta_{k}(t)\right|^{2} + 2 \text{Re}[\lambda_k] \left|\delta_{k}(t)\right|^{2} dt \nonumber \\
     &+ \left( \sum_{l=0}^{n-1} \left[ \left( \frac{4}{n} \sin{\frac{\pi k}{n}} \sin{\frac{\pi l}{n}}\right)^2
     |\delta_l(t)|^2 \right] -\frac{8}{n} \sin^2{\frac{\pi k}{n}} |\delta_k(t)|^2 \right) d L^{pn}_{t}.
\end{align}
Therefore, the time evolution of the power spectrum can be represented more concisely in the form:
\begin{equation}\label{eq:flip_pn}
    d \boldsymbol{P}=2 \operatorname{Re}[\Lambda] \boldsymbol{P} d t+W \boldsymbol{P} d L^{p n}_{t},
\end{equation}
where $\Lambda$ is given in \eqref{Diagonal}, $\boldsymbol{P} = (|\delta_0(t)|^2, |\delta_1(t)|^2, \cdots, |\delta_k(t)|^2, \cdots, |\delta_{n-1}(t)|^2)^\mathsf{T}$ and
\begin{equation}
    \left\{ W \right\}_{l,m} = 
    \begin{cases}
     - \frac{8}{n} \sin^2{\frac{\pi (l-1)}{n}} + \left( \frac{4}{n} \sin^2{\frac{\pi (l-1)}{n}} \right)^2 & (l=m) \\
    \left( \frac{4}{n} \sin{\frac{\pi (l-1)}{n}} \sin{\frac{\pi (m-1)}{n}} \right)^2 \ & (\rm{otherwise}) .
    \end{cases}
\end{equation}

\section{Derivation of the time evolution of the power spectrum (\ref{eq:sigma})}
From equation (\ref{eq:chatj}), the components of the matrix $\hat{C}^j$ are given as below:
\begin{equation}
    \left\{\hat{C}^{j}\right\}_{k, l}=
    \begin{cases}
        \sqrt{(n+1)/n} & (\text{if } k=l=1) \\
        -\frac{1}{\sqrt{n(n+1)}} \frac{\sin \left(\frac{\pi(l-1)}{n}\right)}{\sin \left(\frac{\pi(k-1)}{n+1}-\frac{\pi(l-1)}{n}\right)} 
        e^{i \pi (\frac{(2j-1) (k-1)}{n+1}-\frac{2(j-1) (l-1)}{n})} & \text{(otherwise)}
    \end{cases}.
\end{equation}
The power spectrum after proliferation of cell $j$ is obtained from the Fourier coefficient $\delta_k$ before proliferation as below:
\begin{align}
    {| \delta_{k-1} |^2}^{\ j}_\text{after} &= \left( \sum_{l=1}^n \{\hat{C}^j\}_{k,l} \delta_{l-1} \right) \left( \sum_{m=1}^n \{\overline{\hat{C}}^j_{k,m}\} \overline{\delta}_{m-1} \right) \nonumber \\
    &= \sum_{m=1}^n \sum_{l=1}^n \left[ \{\hat{C}^j\}_{k,l} \{\overline{\hat{C}}^j\}_{k,m} \delta_{l-1} \overline{\delta}_{m-1} \right].
\end{align}
As in the cell mixing model, the time evolution of the power spectrum is approximated by replacing the effect of each proliferation event with an average effect.
Considering the average effect on the power spectrum, we calculate the average of ${| \delta_k |^2}^{\ j}_\text{after}$ over $j$:
\begin{align} \label{delkafter}
    \frac{1}{n} \sum_{j=1}^n {| \delta_{k-1} |^2}^{\ j}_\text{after} &= \frac{1}{n} \sum_{j=1}^n \sum_{m=1}^n \sum_{l=1}^n \left[ \{\hat{C}^j\}_{k,l} \{\overline{\hat{C}}^j\}_{k,m} \delta_{l-1} \overline{\delta}_{m-1} \right] \nonumber \\
    &= \frac{1}{n} \sum_{m=1}^n \sum_{l=1}^n \delta_{l-1} \overline{\delta}_{m-1} \left[ \sum_{j=1}^n \{\hat{C}^j\}_{k,l} \{\overline{\hat{C}}^j\}_{k,m}  \right],
\end{align}
and
\begin{align}
    \sum_{j=1}^n \{\hat{C}^j\}_{k,l} \{\overline{\hat{C}}^j\}_{k,m} =
    \begin{cases}
        0 & \text{if } l \neq m \\
        \frac{1}{(n+1)} \frac{\sin ^{2} \frac{\pi(l-1)}{n}}{\sin ^{2}\left(\frac{\pi(k-1)}{n+1}-\frac{\pi(l-1)}{n}\right)} & \text{if } l=m \text{ and } l \neq 1 \\
        (n+1) & \text{if } k=l=m=1.
    \end{cases}
\end{align}
Here we used the fact that
\begin{equation}
    \sum_{j=1}^n \mathrm{e}^{i2\pi(j-1)(l-m)/n} =
    \begin{cases}
        0 & \text{if } l \neq m \\
        n & \text{if } l = m.
    \end{cases}
\end{equation}
Hence, 
\begin{align}\label{eq:prolav}
    \frac{1}{n} \sum_{j=1}^n {| \delta_{k-1} |^2}^{\ j}_\text{after} =
    \begin{cases}
        \sum_{l=1}^n \frac{1}{n(n+1)} \frac{\sin^2 \left(\frac{\pi(l-1)}{n}\right)}{\sin^2 \left(\frac{\pi(k-1)}{n+1}-\frac{\pi(l-1)}{n}\right)} |\delta_{l-1}|^2 & \text{if } k \neq 1 \\
        \frac{n+1}{n} |\delta_0|^2 + \frac{1}{n(n+1)} \sum_{l=2}^n |\delta_{l-1}|^2 & \text{if } k=1.
    \end{cases}
\end{align}
Therefore, the effect of a single proliferation event on the power spectrum is represented by the matrix $S$ in (\ref{eq:dPprol}).

Since the Delta expression $D_x$ are real values, $P_k = P_{n-k}$ hold.
Because of this symmetry, $\boldsymbol{P}_n$ is represented by the superposition of cosine waves:
\begin{align}
    \boldsymbol{P}_n &= \sum_{k=0}^{n-1} {e}_k \boldsymbol{z}^{n}_k, \\
    \boldsymbol{z}^n_k &= \left(1, \cos{\frac{2\pi k}{n}}, \cos{\frac{4\pi k}{n}}, \cdots, \cos{\frac{2(n-1)\pi k}{n}} \right)^\mathsf{T}.
\end{align}
Here, $e_k$ are the coefficients of superposition.
From the orthogonality of the trigonometric function, we obtain:
\begin{equation}
    \boldsymbol{e} = Z \boldsymbol{P}_n,
\end{equation}
where $\boldsymbol{e} = (e_0, e_1, \cdots, e_{n-1})^\mathsf{T}$ and $Z$ is a square $n\times n$ matrix such that:
\begin{equation}
    \left\{ {Z} \right\}_{l,m} = \cos{\frac{2 \pi (l-1)(m-1)}{n}}.
\end{equation}
From the symmetry of $\boldsymbol{P}_n$, we can also obtain $e_k$ as a discrete Fourier transform of $\boldsymbol{P}_n$.
As the discrete Fourier transform of the power spectrum is the auto-correlation function (from the Wiener-Khinchin theorem), $e_k$ corresponds to the averaged auto-correlation function of $D_x$.

$S\boldsymbol{P}_n$ is also represented by the superposition of cosine waves with different coefficients $\hat{e}_k$:
\begin{align}\label{eq:SPsuperposition}
    S\boldsymbol{P}_n &= \sum_{k=0}^{n} \hat{e}_k \boldsymbol{z}^{n+1}_k.
\end{align}
Therefore, the power spectra $\boldsymbol{P}_n$ and $S\boldsymbol{P}_n$ can be regarded as the sampled values of the function $P(\theta)$ and $P(\theta)_{\text{after}}$, respectively:
\begin{align}
    P(\theta) = \sum_{k=0}^{n-1} e_k \cos{kx} \\
    P(\theta)_{\text{after}} = \sum_{k=0}^{n} \hat{e}_k \cos{kx}.
\end{align}
Then the matrix $S$ can be regarded as a map that transfers the coefficients of superposition $e_k$ to $\hat{e}_k$.

The vector $2\text{Re}[\Lambda] \boldsymbol{P}_n$ is also regarded as the sampled values of the function $2\lambda(\theta) P(\theta)$, where
\begin{equation}
    \lambda(\theta)=\text{Re} \left[ \frac{-(a+d) + \sqrt{(a+d)^{2}-4(a d + 2b \alpha \cos{\theta})}}{2} \right].
\end{equation}
Therefore, the stability of the power spectrum vector $\boldsymbol{P}_n$ can be examined by approximating $S$ with a square matrix $\Sigma$ such that $\Sigma \boldsymbol{P}_n$ share the same coefficients of the superposition with $S\boldsymbol{P}_n$.

We write
\begin{equation}
    S = \frac{1}{n} \hat{Z} Q Z,
\end{equation}
where $\hat{Z}$ is a square $(n+1) \times (n+1)$ matrix and $Q$ is an $(n+1) \times n$ matrix whose components are, respectively,
\begin{equation}
    \left\{ \hat{Z} \right\}_{l,m} = \cos{\frac{2 \pi (l-1)(m-1)}{n+1}},
\end{equation}
and
\begin{equation}
    \left\{ Q \right\}_{l,m} = 
    \begin{cases}
        (n+2-l)/(n+1) & (\text{if } l=m) \\
        (l-1)/(n+1)   & (\text{if } l+1=m \text{ and } l \geq 2) \\
        1/(n+1)       & (\text{if } l=n \text{ and } m = 1) \\
        0             & (\text{otherwise})
    \end{cases}.
\end{equation}
Therefore, the coefficients $\hat{e}_k$ are determined by $e_k$ as follows:
\begin{align}
    \hat{e}_0 &= e_0 \nonumber\\
    \hat{e}_k &= \frac{n+1-k}{n+1} e_k + \frac{k}{n+1} e_{k-1} \quad (1 \leq k \leq n-1) \nonumber \\
    \hat{e}_{n} &= \frac{n}{n+1} e_{n-1} + \frac{1}{n+1} e_{0}.  
\end{align}
This relationship is derived from the formulae in Supplementary text B \cite{SupMat}.

When $n$ is even, we define an $n \times n$ square matrix $\hat{Q}$ by removing the $(n/2+1)$-th row of the matrix $Q$, and then define an $n \times n$ square matrix $\Sigma$ such that:
\begin{equation}
    \Sigma = Z \hat{Q} Z.
\end{equation}
Here,
\begin{align}\displaystyle
    \left\{ \Sigma \right\}_{l,m} = 
    &\frac{2}{n} \sum_{k=2}^{{n}/{2}} \cos \frac{2 \pi(m-1)(k-1)}{n}\left[\frac{k-1}{n+1} \cos \frac{2 \pi(l-1)(k-2)}{n}+\left(1-\frac{k-1}{n+1}\right) \cos \frac{2 \pi(l-1)(k-1)}{n}\right] \nonumber\\
    &+\frac{1}{n}\left(1+(-1)^{m+l-2}\left(1-\frac{n}{n+1} \sin ^{2} \frac{\pi(l-1)}{n}\right)\right).
\end{align}
The $n$ dimensional vector $\Sigma \boldsymbol{P}_n$ is represented as the superposition of the cosine waves:
\begin{equation}\label{eq:SigPsuperposition}
    \Sigma \boldsymbol{P}_n = 
    \sum_{k=0}^{n/2}     \hat{e}_k \boldsymbol{z}^{n}_k 
    + \sum_{k=n/2+1}^{n-1}     \hat{e}_{k+1} \boldsymbol{z}^{n}_k .
\end{equation}
Since $z_k^n = z_{n-k}^n$ holds and $n$ is even, equation \eqref{eq:SPsuperposition} can be simplified:
\begin{align}
    S \boldsymbol{P}_n &=
        \sum_{k=0}^{n/2} \tilde{e}_k \boldsymbol{z}_k^{n+1} \label{eq:SPeven}\\
    \tilde{e}_k &= 
    \begin{cases}
        \hat{e}_k & \text{(if $k = 0$)} \\
        \hat{e}_k + \hat{e}_{n-k} &\text{(otherwise)}
    \end{cases}.
\end{align}
Equation \eqref{eq:SigPsuperposition} can also be simplified:
\begin{equation}\label{eq:SigEven}
    \Sigma \boldsymbol{P}_n = 
    \left( \sum_{k=0}^{n/2-1} \tilde{e}_k \boldsymbol{z}_k^{n} \right) + \hat{e}_{n/2} \ \boldsymbol{z}_{n/2}^{n}.
\end{equation}

When $n$ is odd, we define an $n \times n$ square matrix $\hat{Q}$ by removing the $((n+3)/2)$-th row of the matrix $Q$, and define an $n \times n$ square matrix $\Sigma$ such that:
\begin{equation}
    \Sigma = Z \hat{Q} Z.
\end{equation}
Here,
\begin{align}\displaystyle
    \left\{ \Sigma \right\}_{l,m} = 
    &\frac{2}{n} \sum_{k=2}^{{(n+1)}/{2}} \cos \frac{2 \pi(m-1)(k-1)}{n}\left[\frac{k-1}{n+1} \cos \frac{2 \pi(l-1)(k-2)}{n}+\left(1-\frac{k-1}{n+1}\right) \cos \frac{2 \pi(l-1)(k-1)}{n}\right] \nonumber \\
    &+\frac{1}{n}.
\end{align}
The $n$ dimensional vector $\Sigma \boldsymbol{P}_n$ is represented as the superposition of the cosine waves:
\begin{equation}\label{eq:SigPsuperposition2}
    \Sigma \boldsymbol{P}_n = 
    \sum_{k=0}^{(n-1)/2}  \hat{e}_k \boldsymbol{z}^{n}_k 
    +\sum_{k=(n+1)/2}^{n-1}  \hat{e}_{k+1} \boldsymbol{z}^{n}_k .
\end{equation}
Since $n$ is odd, equation \eqref{eq:SPsuperposition} can be simplified:
\begin{align}
    S \boldsymbol{P}_n &=
    \sum_{k=0}^{(n+1)/2} \tilde{e}_k \boldsymbol{z}_k^{n+1} \label{eq:SPodd}\\
    \tilde{e}_k &= 
    \begin{cases}
        \hat{e}_k & \text{(if $k = 0$ or $(n+1)/2$)} \\
        \hat{e}_k + \hat{e}_{n-k} & \text{(otherwise)}
    \end{cases}.
\end{align}
Equation \eqref{eq:SigPsuperposition2} can also be simplified:
\begin{equation}\label{eq:SigOdd}
    \Sigma \boldsymbol{P}_n = 
        \sum_{k=0}^{(n-1)/2} \tilde{e}_k \boldsymbol{z}_k^{n}.
\end{equation}

Comparing \eqref{eq:SigEven} with \eqref{eq:SPeven} and \eqref{eq:SigOdd} with \eqref{eq:SPodd}, $\Sigma \boldsymbol{P}_n$ and $S \boldsymbol{P}_n$ can be represented by the same cosine wave superposition except for that of the shortest wavelength ($\tilde{e}_{n/2}$ when $n$ is even, $\tilde{e}_{(n+1)/2}$ when $n$ is odd).

The shortest wavelength component of the superposition $\tilde{e}_{n/2}$ or $\tilde{e}_{(n+1)/2}$ corresponds to the long-range correlation of the Delta expression pattern $\boldsymbol{D}_n$.
Since the Delta-Notch interaction and cell proliferation locally affects the pattern, we expect the long-range correlation to be small.
Thus, the contribution of the shortest wavelength component of the cosine wave superposition alone to the spectral structure of the power spectrum would be small when $n$ is sufficiently large.
Therefore, $\Sigma$ is a square matrix that approximates $S$, in the sense that it preserves the spectral structure of the power spectrum.
Based on this assumption, we can analyze equation \eqref{eq:dPprol} in the same way as in the cell mixing model by replacing $S$ with $\Sigma$, and find that it gives the results that agree with the numerical results of the cell proliferation model \eqref{eq:prolModel} (Fig.~\ref{fig:4} (D,F)).

\onecolumngrid
\clearpage

\begin{center}
	\textbf{\large Supplementary material for ``Analyzing the effect of cell rearrangement on Delta-Notch pattern formation'' }\\[5pt]
\end{center}
\setcounter{equation}{0}
\setcounter{table}{0}
\setcounter{page}{1}
\setcounter{section}{0}

\renewcommand{\figurename}{Supplementary Figure}
\renewcommand{\theequation}{S\arabic{equation}}
\renewcommand{\thepage}{S\arabic{page}}
\def\thesection{\Alph{section}}
\setcounter{figure}{0}

In this supplementary material, we provide the proof of the formulae and the additional data of the numerical simulations and the experiment.

\begin{center}
	\textbf{\normalsize A. Approximation of \eqref{eq:flipdP} by a deterministic nonlocal evolution equation}\\[5pt]
\end{center}

The time evolution of the power spectrum \eqref{eq:flipdP} can be approximated by a deterministic nonlocal evolution equation by assuming that the number of cells $n$ is sufficiently large.

Here, the $k$-th component of $d \boldsymbol{P} $ in \eqref{eq:flipdP} is:
\begin{align}
    \left\{ d \boldsymbol{P} \right\}_k
    &= 2 \text{Re}[\lambda_{k-1}] \left\{ \boldsymbol{P} \right\}_k + \left\{W \boldsymbol{P} d L_{t}^{pn}\right\}_k \nonumber \\
    &= 2 \text{Re}[\lambda_{k-1}] P_{k-1} + \left[16 \sin ^{2} \frac{\pi(k-1)}{n}\left(\frac{1}{n} \sum_{l=1}^{n} P_{l-1} \sin ^{2} \frac{\pi (l-1)}{n}\right)-8 \sin ^{2} \frac{\pi(k-1)}{n} P_{k-1}\right] \frac{d L_{t}^{pn}}{n} \nonumber \\
    &= 2\lambda(\theta_{k,n}) P(\theta_{k,n}) + \left[16 \sin ^{2} \frac{\theta_{k,n}}{2}
    \left(\frac{1}{n} \sum_{l=0}^{n-1} P(\varphi_{l,n}) \sin ^{2} \frac{\varphi_{l,n}}{2}\right)
    -8 \sin ^{2} \frac{\theta_{k,n}}{2} P(\theta_{k,n}) \right] \frac{d L_{t}^{pn}}{n}.
\end{align}
Here, we denote $\theta_{k,n} = 2\pi(k-1)/n$ and $\varphi_{l,n} = 2\pi(l-1)/n$ and regard the power spectrum $P_k$ as a function of $\theta_{k,n}$. 
As $n \to \infty$, $\theta_{k,n} \to \theta$ and $\theta$ is dense in $[0,2\pi)$, so that $P(\theta,t)$ becomes a continuous function of $\theta \in [0,2\pi)$ and $t \in [0,\infty)$.
Also, the diagonal matrix $\text{Re}[\Lambda]$ is regarded as a multiplication operator with multiplication factor $2\lambda(\theta)$ where
\begin{equation}\label{lambdatheta}
    \lambda(\theta) = \text{Re} \left[ \frac{-(a+d) + \sqrt{(a+d)^{2}-4(a d + 2b \alpha \cos{\theta}))}}{2} \right].
\end{equation}
Using the fact that the Poisson process ${L}^{pn}_t / n$ converges to $pt$ as $n \to \infty$, so that $d{L}^{pn}_t /n \sim p dt$.
Then, by assuming $n \to \infty$,
\begin{align}
    \theta_{k,n} &\xrightarrow{n\to\infty} \theta \quad (\theta \in [0,2\pi)) \\
    \varphi_{l,n} &\xrightarrow{n\to\infty} \varphi \quad (\varphi \in [0,2\pi)) \\
    \frac{1}{n} \sum_{m=0}^{n-1} P(\varphi_{l,n}) \sin ^{2} \frac{\varphi_{l,n}}{2} &\xrightarrow{n \to \infty} \frac{1}{2\pi} \int_{0}^{2\pi} P(\varphi) \sin^2 \frac{\varphi}{2} d\varphi \\
    \frac{dL^{pn}_t}{n} &\xrightarrow{n \to \infty} p dt.
\end{align}
Therefore, \eqref{eq:flipdP} is approximated by a system of deterministic nonlocal evolution equations as:
\begin{align} \label{eq:yPtheta}
    \frac{d}{d t} P(\theta,t)&=\left(2 \lambda(\theta)-8 p \sin ^{2} \frac{\theta}{2}\right) P(\theta,t) \nonumber \\
    & \quad +\frac{8}{\pi} p \sin ^{2} \frac{\theta}{2} \int_{0}^{2 \pi} P(\varphi,t) \sin ^{2} \frac{\varphi}{2} d \varphi \nonumber \\
    &= \mathcal{Y}_p (P(\theta,t)),
\end{align}
where $\mathcal{Y}_p$ is the operator acting on $P(\theta,t)$.
Therefore, by using the maximum eigenvalue and the corresponding eigenfunction of the operator $\mathcal{Y}_p$, we can derive the expected pattern dynamics.

From the Perron-Frobenius theorem and its extension to the integral operator by Jentzsch \cite{Jentzsch1912}, the eigenvalue $y$ corresponding to the positive eigenfunction $P^*(\theta)>0$ is the maximum eigenvalue of the operator $\mathcal{Y}_p$.
By substituting the eigenvalue $y$ and the corresponding eigenfunction $P^*(\theta)$ of the operator $\mathcal{Y}_p$ into (\ref{eq:yPtheta}), we have:
\begin{align} \label{eq:flipEVec1} \displaystyle
    P^*(\theta) =
    \frac{\frac{8}{\pi} p \sin ^{2} \frac{\theta}{2} \int_{0}^{2 \pi} P^*(\varphi) \sin ^{2} \frac{\varphi}{2} d \varphi}
    {y +8 p \sin ^{2} \frac{\theta}{2} -2 \lambda(\theta)}.
\end{align}
Hence we obtain a recursive relation for $P^*(\theta)$:
\begin{align}\displaystyle
    \int_{0}^{2 \pi} P^*(\theta) \sin ^{2} \frac{\theta}{2} &d \theta \nonumber\\
    = \frac{8}{\pi} &
    \int_{0}^{2 \pi}
    \frac{p \sin ^{4} \frac{\theta}{2} \int_{0}^{2 \pi} P^*(\varphi) \sin ^{2} \frac{\varphi}{2} d \varphi}
    {y +8 p \sin ^{2} \frac{\theta}{2} -2 \lambda(\theta)}
    d \theta.
\end{align}
Assuming $P^*(\theta) > 0 \ (\forall \theta \in [0,2\pi))$, the integral of $P^*(\theta) \sin^2(\theta/2)$ is a constant positive value, and we can obtain the eigenvalue $y$ of the operator $\mathcal{Y}_p$ as the solution of the integral equation:
\begin{equation}\label{eq:flipEVal}
    \pi = \int_{0}^{2\pi} \frac{8p\sin^4(\phi/2)}{y+8p\sin^2(\phi/2)- 2\lambda(\phi)} d\phi,
\end{equation}
and the corresponding eigenfunction $P^*(\theta)$ is:
\begin{equation}\label{eq:FlipEVec2}
    P^*(\theta) = \frac{8p\sin^2(\theta/2)}{y+8p\sin^2(\theta/2)-2\lambda(\theta)}.
\end{equation}

We can also obtain the critical frequency $p^*$ as the solution of the integral equation that is obtained by substituting $y=0$ into (\ref{eq:flipEVal}) as below:
\begin{equation}\label{eq:pstar}
    \pi=\int_{0}^{2 \pi} \frac{4 p^* \sin ^{4}(\phi / 2)}{4 p^* \sin ^{2}(\phi / 2)-\lambda(\phi)} d \phi.
\end{equation}

\begin{center}
	\textbf{\normalsize B. Proof of the trigonometric formulae in Appendix C}\\[5pt]
\end{center}

In this section, we prove the formulae that are used in Appendix C, namely
\begin{align}
    &\sum_{l=0}^{n-1} \frac{1}{n(n+1)} \frac{\sin ^{2} \frac{\pi l}{n}}{\sin ^{2}\left(\frac{\pi k}{n+1}-\frac{\pi l}{n}\right)}=1+\frac{2}{n+1} \cos \frac{2 \pi k}{n+1}. \label{n1sum} \\
    &\sum_{l=0}^{n-1} \frac{1}{n(n+1)} \frac{\sin ^{2} \frac{\pi l}{n}}{\sin ^{2}\left(\frac{\pi k}{n+1}-\frac{\pi l}{n}\right)} \cos \frac{2 \pi m l}{n} = \frac{n-m+1}{n+1} \cos \frac{2 \pi k m}{n+1} +\frac{m+1}{n+1} \cos \frac{2 \pi k(m+1)}{n+1}. \label{n2sum}
\end{align}
Here, $k,l,m$ and $n$ are integers, and $n \geq 2$, $k \in [1,n-1]$ and $m \in [1,n-1]$.

For (\ref{n1sum}), we start from the formulae for $x \neq (\pi l/n)$ ($l$ is an arbitrary integer) that are obtained in \cite{WANG_2007, EJSMONT_2021}:
\begin{align}
    \sum_{l=0}^{n-1} \frac{1}{\sin ^{2}\left(x+\frac{\pi l}{n}\right)}=\frac{n^{2}}{\sin ^{2}(n x)}, \label{eqs8}\\
    \sum_{l=0}^{n-1} \cot \left(x+\frac{\pi l}{n}\right)=n \cot (n x). \label{eqs9}
\end{align}
By substituting $x = -\pi k /(n+1)$ into (\ref{eqs8}) and (\ref{eqs9}), we obtain:
\begin{align}
    &\sum_{l=0}^{n-1} \frac{\sin ^{2} \frac{\pi k}{n+1}}{\sin ^{2}\left(\frac{\pi l}{n}-\frac{\pi k}{n+1}\right)}=n^{2}, \label{eqs10} \\
    &\sum_{l=0}^{n-1} \cot \left(\frac{\pi l}{n}-\frac{\pi k}{n+1}\right)=n \cot \frac{\pi k}{n+1}. \label{eqs11}
\end{align}
Here we used the periodicity of $\sin^2{x} $ and $\cot{x}$ (period length = $\pi$) such that:
\begin{align}
    \sin^2\left(-\frac{n\pi k}{n+1}\right) &= \sin^2\left(-\pi k +\frac{\pi k}{n+1}\right) = \sin^2\left(\frac{\pi k}{n+1}\right), \\
    \cot\left(-\frac{n\pi k}{n+1}\right) &= \cot\left(-\pi k+\frac{\pi k}{n+1} \right) = \cot\left(\frac{\pi k}{n+1}\right).
\end{align}
We consider the finite summation as below:
\begin{align}
    \sum_{l=0}^{n-1} \frac{\sin ^{2} \frac{\pi l}{n}-\sin ^{2} \frac{\pi k}{n+1}}{\sin ^{2}\left(\frac{\pi l}{n}-\frac{\pi k}{n+1}\right)} 
    &=  \sum_{l=0}^{n-1} \frac{\sin \left(\frac{\pi l}{n}+\frac{\pi k}{n+1}\right)}{\sin \left(\frac{\pi l}{n}-\frac{\pi k}{n+1}\right)} \nonumber \\
    &=  \sum_{l=0}^{n-1} \frac{\sin \left(\frac{\pi l}{n}-\frac{\pi k}{n+1}+\frac{2\pi k}{n+1}\right)}{\sin \left(\frac{\pi l}{n}-\frac{\pi k}{n+1}\right)} \nonumber \\
    &= n \cos \frac{2 \pi k}{n+1}+\sin \frac{2 \pi k}{n+1} \sum_{l=0}^{n-1} \cot \left(\frac{\pi l}{n}-\frac{\pi k}{n+1}\right) \nonumber \\
    &= n \left( \cos \frac{2 \pi k}{n+1}+\sin \frac{2 \pi k}{n+1} \cot \frac{\pi k}{n+1}\right) \nonumber \\
    &= n\left(1+2 \cos \frac{2 \pi k}{n+1}\right). \label{eqs12}
\end{align}
Here we used the formula:
\begin{equation}\label{waseki}
    \sin^2{\alpha} - \sin^2{\beta} = \frac{1}{2}(\cos 2\beta - \cos 2\alpha) = \sin(\alpha+\beta)\sin(\alpha-\beta),
\end{equation}
and the trigonometric addition formulae.
From (\ref{eqs10}) and (\ref{eqs12}), we obtain:
\begin{equation}
    \sum_{l=0}^{n-1} \frac{\sin ^{2} \frac{\pi l}{n}}{\sin ^{2}\left(\frac{\pi k}{n+1}-\frac{\pi l}{n}\right)}=n^{2}+n +2n \cos \frac{2 \pi k}{n+1}. \label{final2}
\end{equation}
Therefore, we obtain (\ref{n1sum}) by dividing the both sides of (\ref{final2}) by $n(n+1)$.

For \eqref{n2sum}, we start from the formula:
\begin{equation}
    \begin{aligned}
        \sin ^{2} \frac{\pi l}{n} \cos \frac{2 \pi ml}{n} &=\frac{1}{2} \left(1-\cos \frac{2\pi l}{n}\right) \cos \frac{2 \pi ml}{n} \\
        &=-\frac{1}{4} \cos  \frac{2\pi(m-1) l}{n}+\frac{1}{2} \cos  \frac{2 \pi ml}{n}-\frac{1}{4} \cos \frac{2 \pi (m+1) l}{n}
    \end{aligned},
\end{equation}
and use it to transform the right-hand of \eqref{n2sum} as follows:
\begin{equation} \label{eq:s14D}
   \frac{1}{n+1} \left( \frac{1}{2} \mathcal{D}_{m}-\frac{1}{4} \mathcal{D}_{m-1}-\frac{1}{4} \mathcal{D}_{m+1} \right),
\end{equation}
where
\begin{equation}
    \mathcal{D}_{m}=\frac{1}{n} \sum_{l=0}^{n-1} \frac{\cos \frac{ 2 \pi m l}{n}}{\sin ^{2}(\frac{\pi k}{n+1}-\frac{\pi l}{n})}.
\end{equation}

We consider the finite summation as below:
\begin{align}
    \mathcal{A}_{m}
    =&\frac{1}{n} \sum_{l=0}^{n-1} \frac{\cos \frac{2\pi mk}{n+1}-\cos \frac{2\pi ml}{n}}{\sin ^{2}(\frac{\pi k}{n+1}-\frac{\pi l}{n})} \nonumber\\
    =&2\cos \frac{2\pi mk}{n+1}\left(\frac{1}{n} \sum_{l=0}^{n-1} \frac{\sin ^{2} m(\frac{\pi k}{n+1}-\frac{\pi l}{n})}{\sin ^{2}(\frac{\pi k}{n+1}- \frac{\pi l}{n})}\right)
    -2\sin \frac{2\pi mk}{n+1}\left(\frac{1}{n} \sum_{l=0}^{n-1} \frac{\sin m(\frac{\pi k}{n+1}-\frac{\pi l}{n}) \cos m(\frac{\pi k}{n+1}-\frac{\pi l}{n})}{\sin ^{2}(\frac{\pi k}{n+1}-\frac{\pi l}{n})}\right) \nonumber \\
    =&2\cos \frac{2\pi mk}{n+1} \mathcal{B}_m
    -2\sin \frac{2\pi mk}{n+1} \mathcal{C}_m \label{mathcalA}.
\end{align}
Here we used the formula:
\begin{align}
\cos \frac{2\pi mk}{n+1}-\cos \frac{2\pi ml}{n} 
=&-2 \sin m \left(\frac{\pi k}{n+1}+\frac{\pi l}{n} \right) \sin m \left(\frac{\pi k}{n+1}-\frac{\pi l}{n} \right) \nonumber\\
=&2\left(\sin ^{2} m \left(\frac{\pi k}{n+1}-\frac{\pi l}{n} \right) \cos \frac{2\pi mk}{n+1} \right. \nonumber\\
& \quad \left. -\cos m \left(\frac{\pi k}{n+1} -\frac{\pi l}{n} \right) \sin m \left( \frac{\pi k}{n+1}-\frac{\pi l}{n} \right) \sin \frac{2\pi mk}{n+1}\right),
\end{align}
and define
\begin{align}
    \mathcal{B}_{m}&=\frac{1}{n} \sum_{l=0}^{n-1} \frac{\sin ^{2}  m(\frac{\pi k}{n+1}-\frac{\pi l}{n})}{\sin ^{2}(\frac{\pi k}{n+1}-\frac{\pi l}{n})}, \label{mathcalB}\\
    \mathcal{C}_{m}&=\frac{1}{n} \sum_{l=0}^{n-1} \frac{\sin m(\frac{\pi k}{n+1}-\frac{\pi l}{n}) \cos m(\frac{\pi k}{n+1}-\frac{\pi l}{n})}{\sin ^{2}(\frac{\pi k}{n+1}-\frac{\pi l}{n})}. \label{mathcalC}
\end{align}

For \eqref{mathcalB}, we consider $\Delta \mathcal{B}_{m}$ as a difference between $\mathcal{B}_{m+1}$ and $\mathcal{B}_m$:
\begin{align}
    \Delta \mathcal{B}_{m}
    &=\mathcal{B}_{m+1}-\mathcal{B}_{m} \nonumber \\
    &=\frac{1}{2 n} \sum_{l=0}^{n-1} \frac{\cos \left[ 2 m(\frac{\pi k}{n+1}-\frac{\pi l}{n}) \right] -\cos \left[ 2(m+1)(\frac{\pi k}{n+1}-\frac{\pi l}{n}) \right]}{\sin ^{2}(\frac{\pi k}{n+1}-\frac{\pi l}{n})} \nonumber \\
    &=\frac{1}{2 n} \sum_{l=0}^{n-1} \frac{2 \sin \left[ (2 m+1)(\frac{\pi k}{n+1}-\frac{\pi l}{n}) \right] \sin (\frac{\pi k}{n+1}-\frac{\pi l}{n})}{\sin ^{2}(\frac{\pi k}{n+1}-\frac{\pi l}{n})} \nonumber \\
    &=\frac{1}{n} \sum_{l=0}^{n-1} \frac{\sin \left[(2 m+1)(\frac{\pi k}{n+1}-\frac{\pi l}{n})\right] }{\sin (\frac{\pi k}{n+1}-\frac{\pi l}{n})}.
\end{align}
Here we used the formula:
\begin{equation}
    \sin ^{2} \left[ m \left(\frac{\pi k}{n+1}-\frac{\pi l}{n} \right) \right]=\frac{1-\cos \left[ 2 m(\frac{\pi k}{n+1}-\frac{\pi l}{n}) \right]}{2}.
\end{equation}
We also obtain the difference between $\Delta \mathcal{B}_{m+1}$ and $\Delta \mathcal{B}_m$:
\begin{align}
    \Delta \mathcal{B}_{m+1}-\Delta \mathcal{B}_{m} &=\frac{1}{n}  \sum_{l=0}^{n-1} \frac{\sin \left[(2 m+3)(\frac{\pi k}{n+1}-\frac{\pi l}{n}) \right]-\sin \left[ (2 m+1)(\frac{\pi k}{n+1}-\frac{\pi l}{n}) \right]}{\sin (\frac{\pi k}{n+1}-\frac{\pi l}{n})} \nonumber \\
    &=\frac{1}{n} \sum_{l=0}^{n-1} \frac{2 \cos \left[ 2(m+1)(\frac{\pi k}{n+1}-\frac{\pi l}{n}) \right] \sin (\frac{\pi k}{n+1}-\frac{\pi l}{n})}{\sin (\frac{\pi k}{n+1}-\frac{\pi l}{n})} \nonumber \\
    &=\frac{1}{n}  \sum_{l=0}^{n-1} 2 \cos \left[ 2(m+1) \left(\frac{\pi k}{n+1}-\frac{\pi l}{n} \right) \right] \nonumber \\
    &=0. \label{difBconst}
\end{align}
The relationship \eqref{difBconst} holds for any $m \in \mathbb{N}$.
This means that $\Delta \mathcal{B}_m$ takes the same value independent of $m$:
\begin{equation}
    \Delta \mathcal{B}_{m}=\Delta \mathcal{B}_{0}=\frac{1}{n} \sum_{\kappa=0}^{n-1} 1=1.
\end{equation}
Since $\mathcal{B}_1 = 1$, we obtain:
\begin{equation} \label{mathcalBm}
    \mathcal{B}_{m}= \mathcal{B}_1 + \sum_{\kappa=1}^{m-1} 1 = m.
\end{equation} 

For \eqref{mathcalC}, we define $\Delta \mathcal{C}_{m}$ as the difference between $\mathcal{C}_{m+1}$ and $\mathcal{C}_m$:
\begin{align}
    \Delta \mathcal{C}_{m}&=\mathcal{C}_{m+1}-\mathcal{C}_{m} \nonumber \\
    &=\frac{1}{2 n} \sum_{l=0}^{n-1} \frac{\sin \left[ 2 m(\frac{\pi k}{n+1}-\frac{\pi l}{n}) \right] -\sin \left[ 2(m+1)(\frac{\pi k}{n+1}-\frac{\pi l}{n}) \right]}{\sin ^{2}(\frac{\pi k}{n+1}-\frac{\pi l}{n})} \nonumber \\
    &=\frac{1}{2 n} \sum_{l=0}^{n-1} \frac{2 \cos \left[ (2 m+1)(\frac{\pi k}{n+1}-\frac{\pi l}{n}) \right] \sin (\frac{\pi k}{n+1}-\frac{\pi l}{n})}{\sin ^{2}(\frac{\pi k}{n+1}-\frac{\pi l}{n})} \nonumber \\
    &=\frac{1}{n} \sum_{l=0}^{n-1} \frac{\cos \left[ (2 m+1)(\frac{\pi k}{n+1}-\frac{\pi l}{n}) \right]}{\sin (\frac{\pi k}{n+1}-\frac{\pi l}{n})}.
\end{align}
Here we used the formula:
\begin{equation}
    \sin \left[ m \left(\frac{\pi k}{n+1}-\frac{\pi l}{n} \right) \right] \cos \left[ m \left(\frac{\pi k}{n+1}-\frac{\pi l}{n} \right) \right] =\frac{1}{2} \sin \left[ 2 m \left(\frac{\pi k}{n+1}- \frac{\pi l}{n} \right) \right].
\end{equation}
We also obtain the difference between $\Delta \mathcal{C}_{m+1}$ and $\Delta \mathcal{C}_m$:
\begin{align}
    \Delta \mathcal{C}_{m+1}-\Delta \mathcal{C}_{m} &=\frac{1}{n}  \sum_{l=0}^{n-1} \frac{\cos \left[ (2 m+3)(\frac{\pi k}{n+1}-\frac{\pi l}{n}) \right] -\cos \left[ (2 m+1)(\frac{\pi k}{n+1}-\frac{\pi l}{n}) \right]}{\sin (\frac{\pi k}{n+1}-\frac{\pi l}{n})} \nonumber \\
    &=\frac{1}{n} \sum_{l=0}^{n-1} \frac{-2 \sin \left[ 2(m+1)(\frac{\pi k}{n+1}-\frac{\pi l}{n}) \right] \sin (\frac{\pi k}{n+1}-\frac{\pi l}{n})}{\sin (\frac{\pi k}{n+1}-\frac{\pi l}{n})} \nonumber \\
    &=\frac{1}{n}  \sum_{l=0}^{n-1} -2 \sin \left[ 2(m+1) \left(\frac{\pi k}{n+1}-\frac{\pi l}{n} \right) \right] \nonumber \\
    &=0. \label{difCconst}
\end{align}
The relationship \eqref{difCconst} holds for any $m \in \mathbb{N}$.
Hence $\Delta \mathcal{C}_m$ takes the same value independent of $m$:
\begin{equation}
    \Delta \mathcal{C}_{m}=\Delta \mathcal{C}_{0}=\frac{1}{n} \sum_{l=0}^{n-1} \frac{\cos (\frac{\pi k}{n+1}-\frac{\pi l}{n})}{\sin (\frac{\pi k}{n+1}-\frac{\pi l}{n})}=\cot \frac{\pi k}{n+1}.
\end{equation}
Here we used equation \eqref{eqs11}.
Since $\mathcal{C}_1 = \cot (\pi k/(n+1))$, we obtain:
\begin{equation}\label{mathcalCm}
    \mathcal{C}_{m}=m  \cot \frac{\pi k}{n+1}.
\end{equation}

By substituting \eqref{mathcalBm} and \eqref{mathcalCm} into \eqref{mathcalA}, we obtain:
\begin{align}\label{mathcalAm}
    \mathcal{A}_m
    &=2 m \cos \frac{2\pi mk}{n+1}-2 m \cot \frac{\pi k}{n+1} \sin \frac{2\pi mk}{n+1} \nonumber \\
    &=-2 m \frac{\sin \frac{(2 m+1)\pi k}{n+1}}{\sin \frac{\pi k}{n+1}} \nonumber \\
    &= -2 m -4 m \left(\cos \frac{2\pi k}{n+1}+\cos \frac{4\pi k}{n+1}+\cdots+\cos \frac{2\pi mk}{n+1} \right).
\end{align}
Here we used the relationship:
\begin{equation}
    \left(\sum_{\kappa=1}^{m} \cos \frac{2 \kappa \pi k}{n+1}\right)  \sin \frac{\pi k}{n+1}=\frac{1}{2} \left( \sin \frac{(2 m+1)\pi k}{n+1}-\sin \frac{\pi k}{n+1} \right),
\end{equation}
which is derived from the formula:
\begin{equation}
    \cos \frac{2\pi mk}{n+1} \sin \frac{\pi k}{n+1}=\frac{1}{2} \left(\sin \frac{(2 m+1)\pi k}{n+1}-\sin  \frac{(2 m-1)\pi k}{n+1} \right).
\end{equation}

From \eqref{mathcalAm} and \eqref{eqs10}, we obtain:
\begin{equation}
    \mathcal{D}_{m}=\frac{n}{\sin ^{2} \frac{\pi k}{n+1}} \cos \frac{2\pi mk}{n+1}-\left(2 m+4 m\left(\cos \frac{2\pi k}{n+1}+\cos \frac{4\pi k}{n+1}+\cdots+\cos \frac{2\pi mk}{n+1}\right)\right).
\end{equation}
Therefore, we have:
\begin{align}\label{ddif}
    \frac{1}{2} D_{m}-\frac{1}{4} D_{m-1}-\frac{1}{4} D_{m+1} =&-(m-1) \cos \frac{2\pi mk}{n+1}+(m+1) \cos \frac{2(m+1)\pi k}{n+1} \nonumber \\
    & \quad +\frac{n}{\sin ^{2} \frac{\pi k}{n+1}}  \sin ^{2} \frac{\pi k}{n+1}  \cos \frac{2\pi mk}{n+1} \nonumber \\
    =&(n-m+1)  \cos \frac{2\pi mk}{n+1}+(m+1) \cos \frac{2(m+1)\pi k}{n+1}.
\end{align}
Here we used the formula:
\begin{equation}
    2\cos \frac{2m \pi k}{n+1} - \cos \frac{2(m-1) \pi k}{n+1} -\cos \frac{2(m+1) \pi k}{n+1} = 4 \sin^2 \frac{\pi k}{n+1} \cos \frac{2m \pi k}{n+1}.
\end{equation}
Therefore, we obtain \eqref{n2sum} by dividing the both sides of \eqref{ddif} by $(n+1)$.

\begin{figure*}[htb]
    \includegraphics[width=17.8cm]{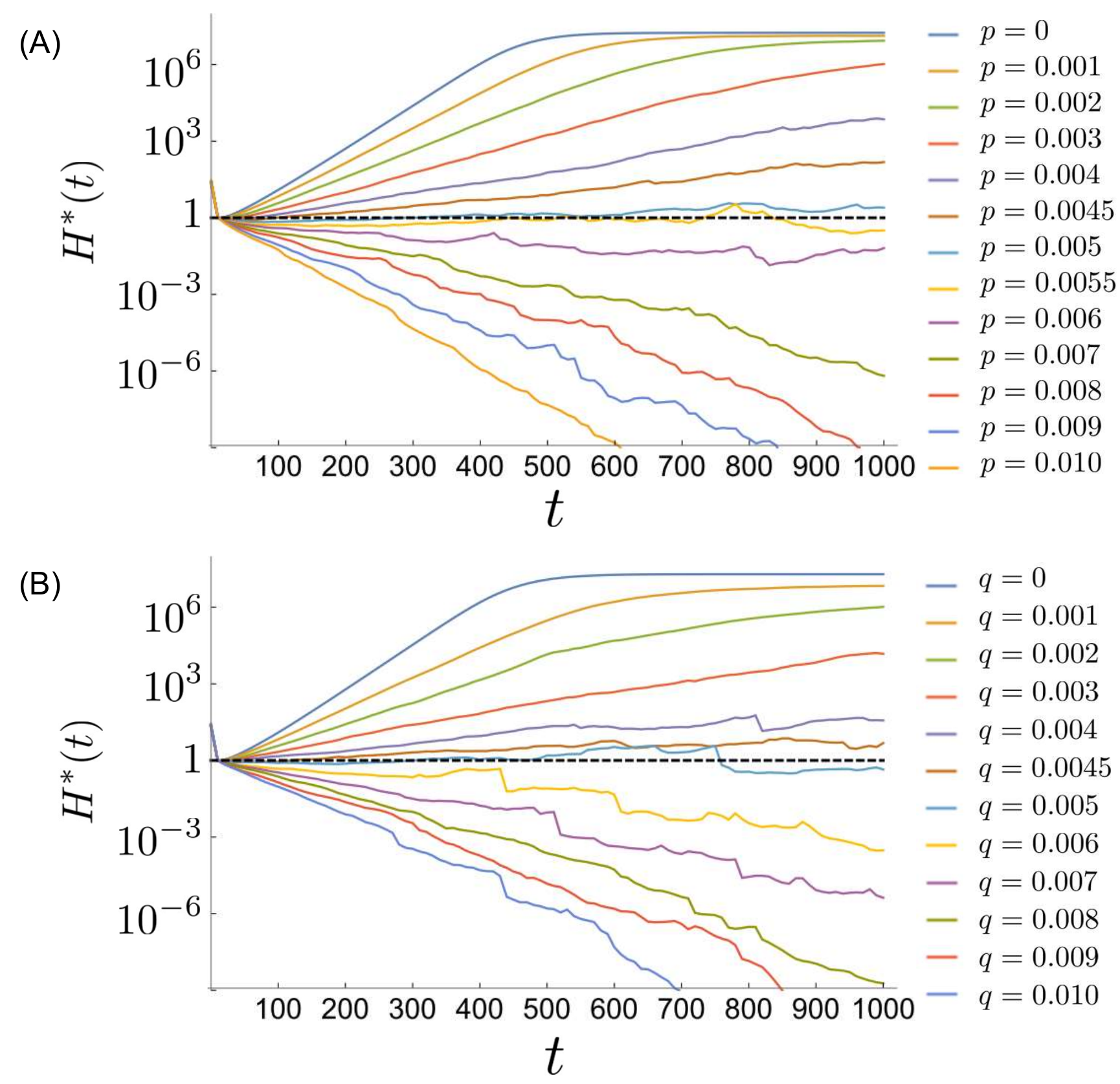}
    \caption{
        Plots of the normalized heterogeneity of the pattern $H^*(t)$ against time $t$.
        The heterogeneity $H^*(t)$ shown in this figure was obtained by averaging the value of $H(t)$ of $400$ different simulations, and normalized by divided by $H_0 = H(10)$ for each $p$ and $q$.
        (A) Time evolution of $H^*(t)$ in the cell mixing model.
        It can be seen that $H^*(t)$ increases for $p \leq 0.0045$ and decreases for $p \geq 0.006$.
        (B) Time evolution of the $H^*(t)$ in the cell proliferation model. 
        It can be seen that $H^*(t)$ increases for $q \leq 0.004$ and decreases for $q \geq 0.006$.
        Other parameter values and initial conditions are as in Fig.~\ref{fig:2}.
        }
    \label{s1}
\end{figure*}

\newpage

\begin{figure*}[htb]
    \includegraphics[width=17.8cm]{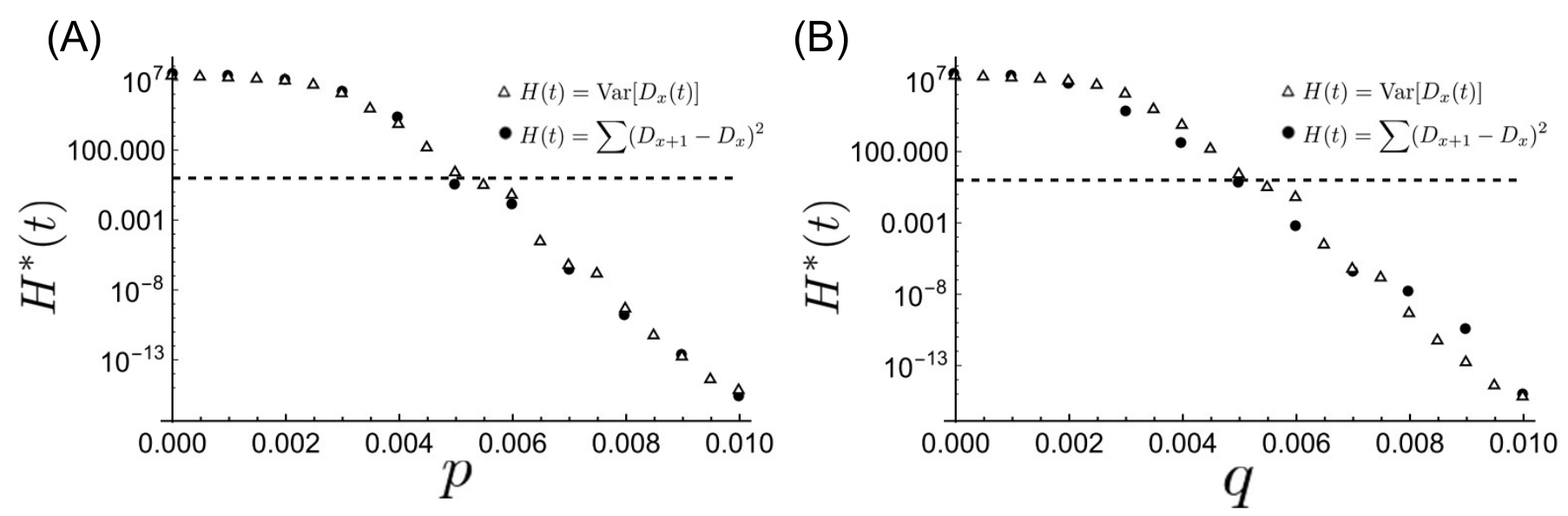}
    \caption{
        Comparison of the different definitions of heterogeneity.
        The triangles represent $H(t)$ that is defined as the variance of $D_x$. The black dots represent the $H(t)$ that is defined as the average of $(D_{x+1}-D_x)^2$ in the cell mixing model (A) and the cell proliferation model (B).
        The heterogeneity function $H^*(t)$ shown in this figure was obtained by averaging the $H(t)$ value of $400$ different simulations, and normalized by dividing by $H_0 = H(10)$ for each $p$ and $q$ in both definitions of heterogeneity.
        The triangles are the values shown in Fig.~\ref{fig:3} for $t=1000$. 
        For the black dots, we calculated the heterogeneity at $11$ different frequencies of $p$ and $q$, which are taken in the range $0$ to $0.01$ at equal intervals of $0.001$ in each model.
        The conditions and parameter values are as in Fig.~\ref{fig:2}.
        }
    \label{s5}
\end{figure*}

\begin{figure*}[htb]
    \includegraphics[width=8.6cm]{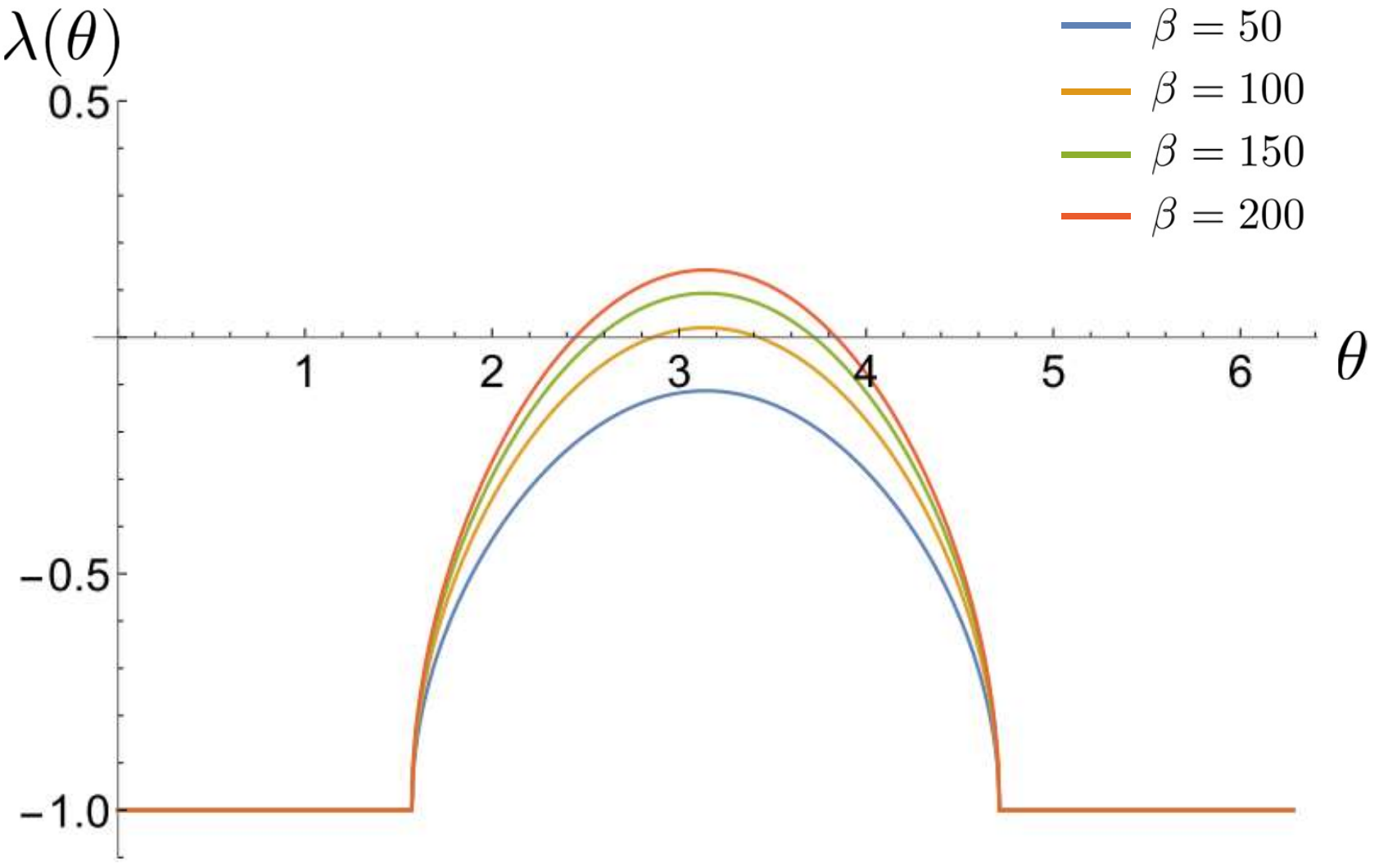}
    \caption{
        Plots of the dispersion-relation for the version of the Collier model we use (\ref{eq:Collier}).
        $\lambda(\theta)$ is the real part of the larger eigenvalue in (\ref{eq:lambdak2}) and $\theta = 2 \pi k/n$. 
        For $\beta=50$, $\lambda(\theta)$ is negative for any $k$.
        For $\beta=100$, $\lambda(\theta)$ takes positive values in a narrow region around $\theta = \pi$.
        As $\beta$ increases, the region where $\lambda(\theta)$ takes positive value increases.
        The other parameters are as in Fig.~\ref{fig:2}.
    }
    \label{s2}
\end{figure*}

\newpage

\begin{figure*}[htb]
    \includegraphics[width=17.8cm]{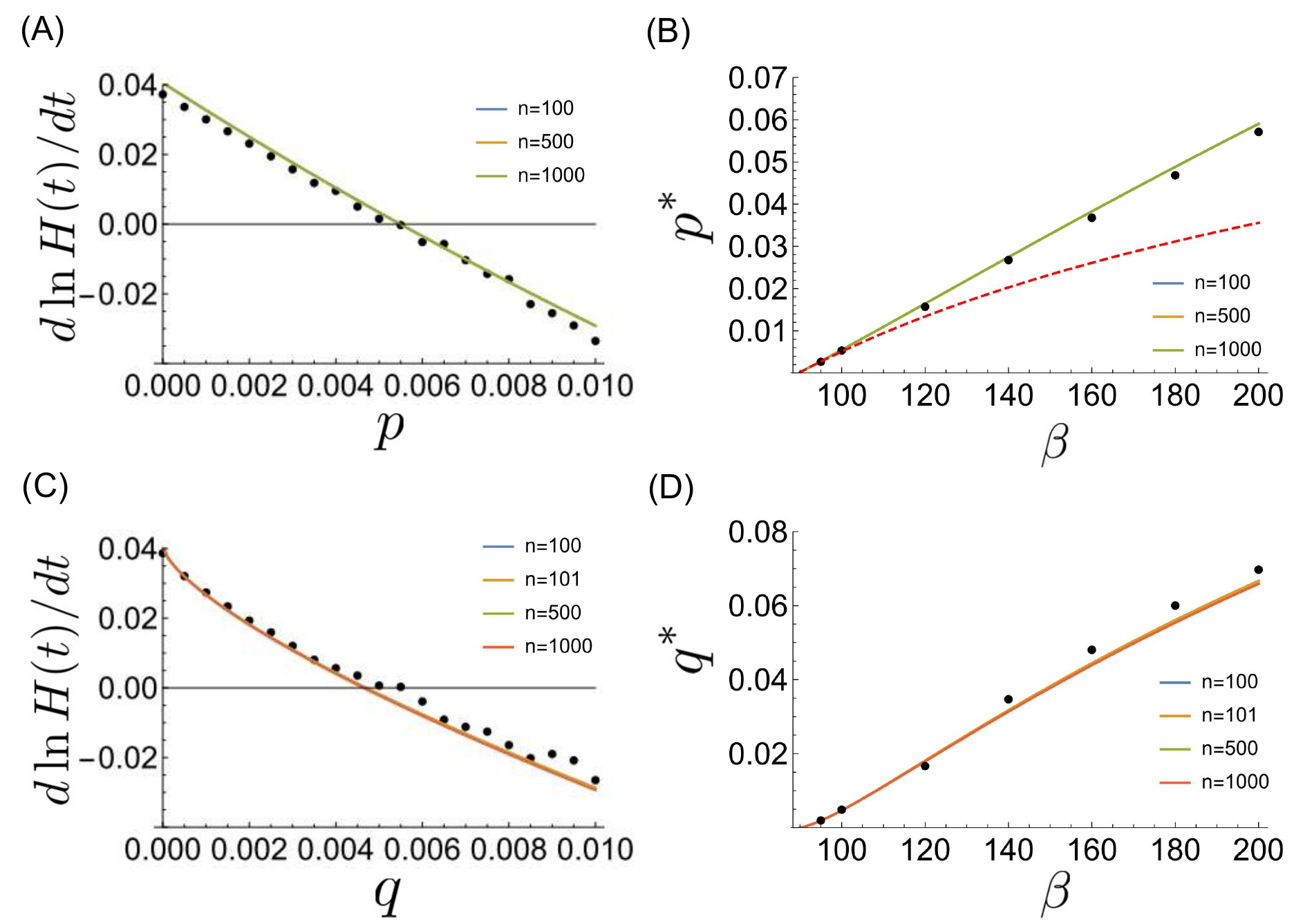}
    \caption{
        Plots of the analytical results at different cell numbers.
        (A) The solid lines represent the maximum eigenvalue $y$ derived from the matrix $Y_p$ \eqref{eq:2lpnw} at different cell numbers ($n=100, 500, 1000$) and the black dots are the growth rate $d \ln{H(t)}/dt$ estimated from Fig.~S1(A), corresponding to Fig.~\ref{fig:4}(A).
        (B) The solid lines represent the values of $p^*$ derived from the \eqref{eq:2lpnw} at different cell numbers ($n=100, 500, 1000$), the red dashed line represents $p^*$ derived from (\ref{eq:maxlk}) and the black dots represent the values of $p^*$ that were estimated from Fig.~\ref{fig:3}(A), corresponding to Fig.~\ref{fig:4}(C).
        (C) The solid lines represent the maximum eigenvalue derived from the matrix $J_q$ \eqref{eq:matJ} at different cell numbers ($n=100, 101, 500, 1000$) and the black dots are the growth rate $d \ln{H(t)}/dt$ estimated from Fig.~S1(B), corresponding to Fig.~\ref{fig:4}(D).
        (D) The solid lines represent the values of $q^*$ derived from the matrix $J_q$ \eqref{eq:matJ} at different cell numbers ($n=100, 101, 500, 1000$), and the black dots represent the values of $q^*$ that were estimated from Fig.~\ref{fig:3}(B), corresponding to Fig.~\ref{fig:4}(F).
        }.
    \label{s6}
\end{figure*}

\newpage

\begin{figure*}[htb]
    \includegraphics[width=17.8cm]{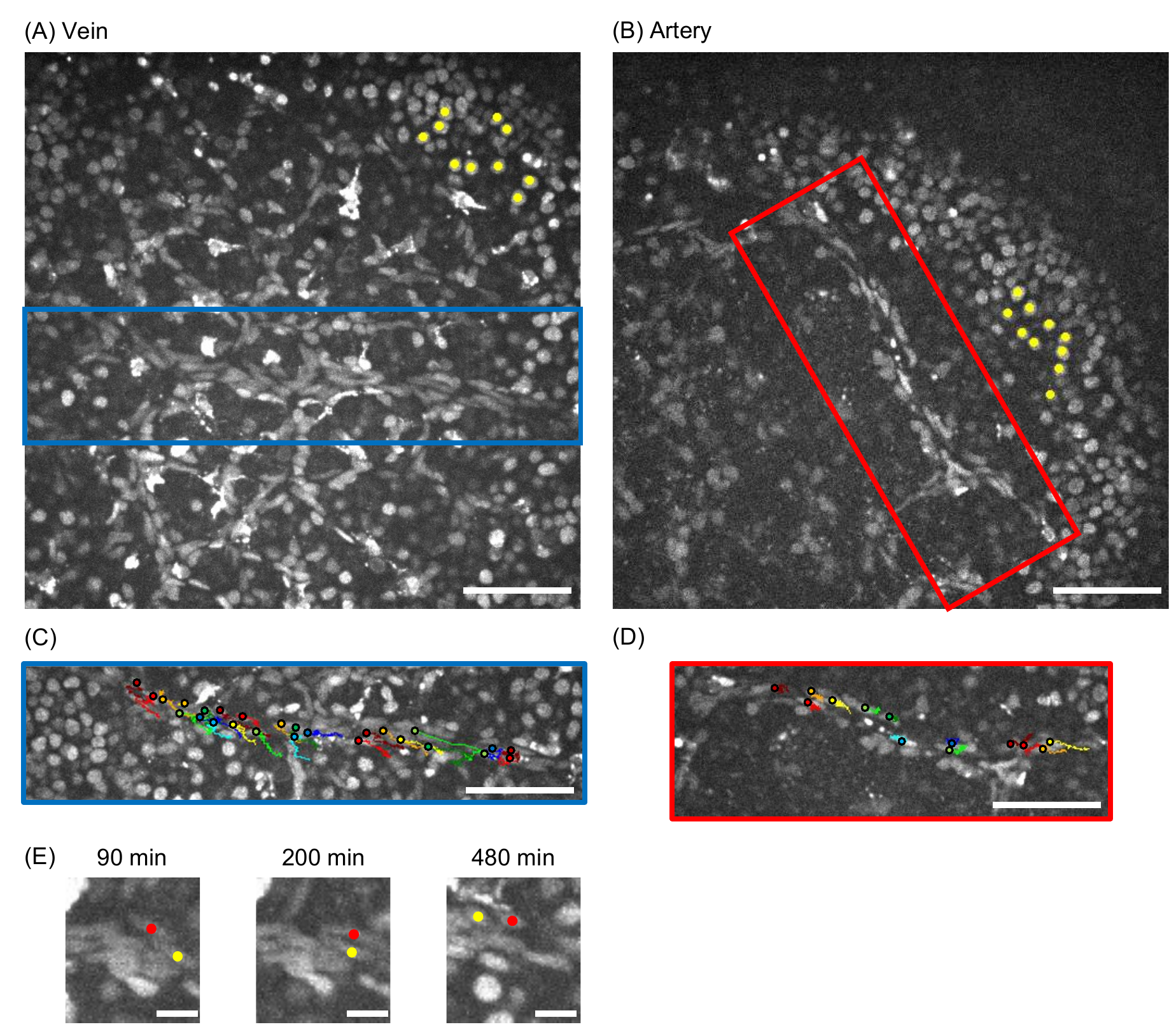}
    \caption{
        Timelapse observations of endothelial cells in the retina.
        (A--B) Microscopic images of the retina.
        The blue rectangle indicates a vein, the red rectangle indicates an artery and the yellow dots indicate the ganglion cells in the retina.
        (C--D) Tracks of endothelial cell movement in vein (C) and artery (D).
        The lines indicate the path of individual endothelial cells and the dots indicate the position of the cells at $t=720$ min.
        (E) Crossing of endothelial cells in the vein.
        The red and yellow dots indicate the endothelial cells. 
        They exchanged their relative position.
        The images (A--D) are taken at $t = 720$ min. 
        Scale bar = $100$ {\textmu}m in (A--D) and $20$ {\textmu}m in (E).
    }
    \label{s3}
\end{figure*}

\newpage

\begin{figure*}[htb]
    \includegraphics[width=17.8cm]{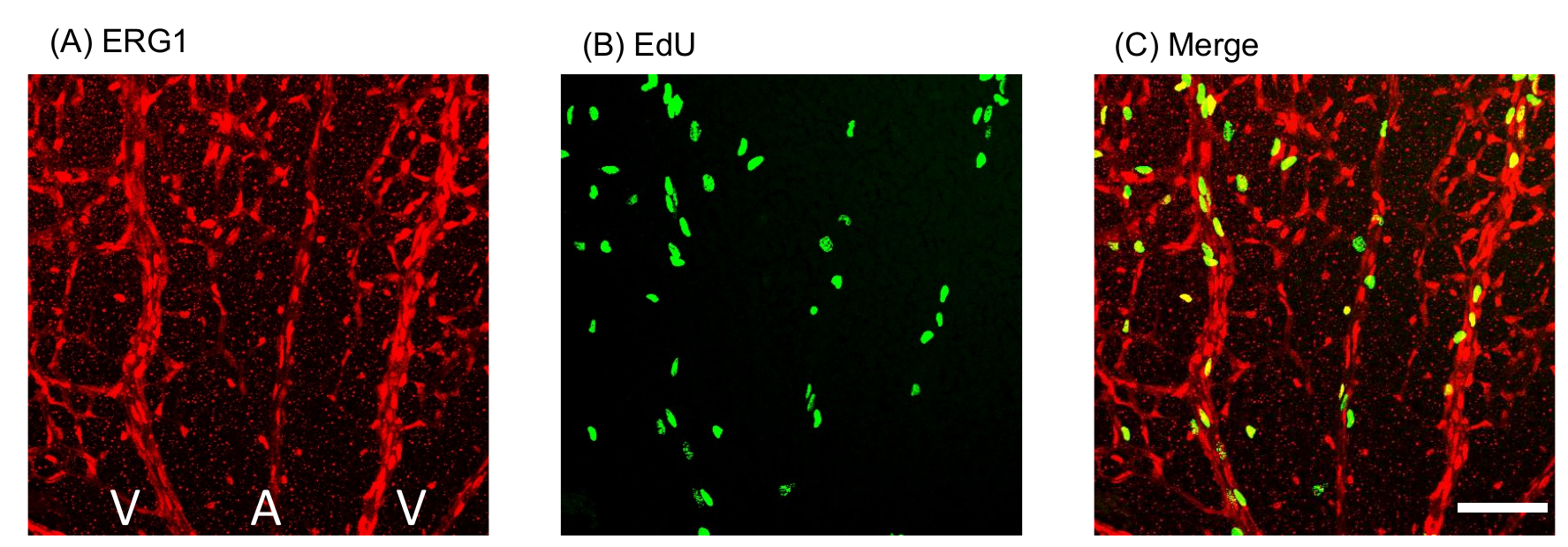}
    \caption{
        Fluorescent images of the staining for ERG1 (A) and EdU (B) in the retinal vasculature.
        ``V'' indicates veins, and ``A'' indicates arteries.
        The vessel type was distinguished by anatomical features such as the avascular zone and the thickness of the vessel.
        EdU was intraperitoneally injected two hours before sacrifice.
        Scale bar = $100$ {\textmu}m}.
    \label{s4}
\end{figure*}

\end{document}